\documentclass[12pt]{article}

\pdfoutput=1

\usepackage{graphics}
\usepackage{amssymb}
\usepackage{amsmath}
\usepackage{tikz}

\newcommand{\beq}{\begin{equation}}

\newcommand{\eeq}{\end{equation}}
\newcommand{\bea}{\begin{eqnarray}}
\newcommand{\eea}{\end{eqnarray}}

\newcommand{\R}{{\rm I\!R}}

\begin{document}

\begin{center}
${}$\\
\vspace{60pt}
{ \Large \bf Introducing Quantum Ricci Curvature
}

\vspace{46pt}

{\sl N. Klitgaard}$\,^{a}$
and {\sl R. Loll}$\,^{a,b}$

\vspace{24pt}
{\footnotesize

$^a$~Institute for Mathematics, Astrophysics and Particle Physics, Radboud University \\ 
Heyendaalseweg 135, 6525 AJ Nijmegen, The Netherlands.\\ 
{email: n.klitgaard@science.ru.nl, r.loll@science.ru.nl}\\

\vspace{10pt}

$^b$~Perimeter Institute for Theoretical Physics,\\
31 Caroline St N, Waterloo, Ontario N2L 2Y5, Canada.\\
{email: rloll@perimeterinstitute.ca}

}
\vspace{48pt}

\end{center}

\begin{center}
{\bf Abstract}
\end{center}

\noindent  Motivated by the search for geometric observables in nonperturbative quantum gravity,
we define a notion of coarse-grained Ricci curvature.
It is based on a particular way of extracting the local Ricci curvature of a smooth Riemannian manifold
by comparing the distance between pairs of spheres with that of their centres. 
The quantum Ricci curvature is designed for use on non-smooth and discrete metric spaces,  
and to satisfy the key criteria of scalability and computability. We test the prescription on a variety
of regular and random piecewise flat spaces, mostly in two dimensions.
This enables us to quantify its behaviour for short lattices distances and
compare its large-scale behaviour with that of constantly curved model spaces. 
On the triangulated spaces considered, the quantum Ricci curvature has good averaging properties
and reproduces classical characteristics on scales large compared to the discretization scale.

\vspace{12pt}
\noindent

\newpage

\section{The case for quantum observables}
\label{intro:sec}

A crucial ingredient for understanding the physics of nonperturbative quantum gravity are observables
that capture the properties of spacetime in a diffeomor\-phism-invariant and background-independent manner, 
all the way down to the Planck scale. Thus far, there are very few quantities we know of that fit the bill.

The nonperturbative path-integral approach of Causal Dynamical Triangulations (CDT) has proven a fertile 
ground for defining and studying such observables \cite{physrep}. 
Firstly, the manifest coordinate-invariance of the underlying piecewise flat Regge geometries 
(the ``triangulations") makes this approach purely geometric, in keeping with the spirit of Einstein's classical theory. 
By contrast, quantum formulations that use a differentiable manifold or some regularized
version thereof as part of their background structure usually require a choice of coordinates to do explicit computations.  
In this case, the implementation of a suitable gauge fixing and the consistent treatment of diffeomorphism symmetry 
in the quantum theory often face protracted difficulties. Secondly, CDT provides a 
well-defined computational framework, in which the expectation values of observables can be measured
and studied systematically in the limit as the UV-cutoff -- the so-called ``lattice spacing" $a$ -- is removed. 

This has enabled the operational definition and quantitative evaluation of several interesting observables. 
Among them, the spectral dimension \cite{spectral} is perhaps the best known. It only requires the existence of an
operator of Laplace-type and is therefore relatively straightforward to implement in a variety of ways also in other quantum gravity 
approaches (see \cite{ReuterSauer,carlip} and references therein). 
Measurements of the spectral dimension of the quantum geometry generated in CDT
quantum gravity first exhibited the phenomenon of ``dynamical dimensional reduction" of spacetime from its classical
value of 4 on macroscopic scales to a value compatible with 2 on the Planck scale. Other observables used to
characterize the micro- and macroscopic properties of quantum spacetime are its Hausdorff dimension and
the so-called volume profile of the universe, i.e. its three-volume as a function of cosmological 
proper time \cite{hausdorffvolume}. The latter has also been investigated recently with the help of functional renormalization 
group techniques \cite{FSPC}.

It would clearly be desirable to have more observables that characterize some analogue of local geometry in the 
Planckian regime, beyond the notions of generalized dimensions currently in use. 
Note that these dimensions have a number of nice properties that we may want other observables to possess also.
They can be defined operationally through length and volume measurements, which do not require the presence of 
a smooth metric $g_{\mu\nu}(x)$, but can be performed on piecewise flat manifolds and even more general
metric spaces. At the same time, they {\it can} also be implemented on a smooth $D$-dimensional metric manifold to determine its 
local dimension, in which case they simply reproduce the value $D$ of its topological dimension. In other words, the
generalized dimensions possess a
well-defined classical limit, which justifies calling them ``dimension" in the first place, even when using them in a non-classical, 
non-smooth context.

Another property of the generalized dimensions is that they can be scaled, in the sense that one and the same measuring prescription
can be implemented at various length scales to extract an ``effective dimension" associated with that scale. For example, one may
obtain a Hausdorff dimension $D_H$ of some metric space by measuring the leading-order behaviour of the volumes $V(r)$ of 
geodesic balls of radius $r$ as a function of $r$. The exponent $D_H$ extracted from a power law of the form
$V(r)\propto r^{D_H}$ will then in general depend on the scale $r$. Of course, care is required when interpreting ``dimension" away
from a regime where it behaves classically. Note also that by using the word ``observable" we do not mean to imply any link
to concrete quantum gravity phenomenology but only a quantity that is operationally well defined in a nonperturbative context. 
Lastly, when working in the context of dynamical triangulations, the usual logic of a ``lattice" regularization\footnote{We put
``lattice" in inverted commas, because the notion is potentially misleading in the context of piecewise flat spaces. In such a formulation,
space(-time) itself is not a lattice, but perfectly continuous. Nevertheless, since CDT works with a small number of standardized
simplicial building blocks, it is natural to measure lengths only along edges and in integer multiples of a unit edge length,
as a convenient approximation.} applies: measurements 
at or near the cut-off scale $a$ are discarded because of their dependence on the details of the regularization, like the
shape of the elementary building blocks. Furthermore, we are only interested in continuum properties, that is, properties that
persist in the limit where the regulators are removed. Most quantities one can define on the lattice will have no continuum analogue, 
and will not exhibit characteristic scaling behaviour in the limit as $a\rightarrow 0$ that would allow us to identify them with
dimensionful, physically interesting continuum quantities.\footnote{The lattice spacing $a$ has the dimension of length.
In what follows, we will for simplicity often work with dimensionless lattice units, which amounts to ``setting $a$ equal to 1".}
 
In the present work, we will introduce a new geometric observable that has many of the desirable properties just described,
a quasi-local quantity we call the ``quantum Ricci curvature".
It is defined in purely geometric terms, without invoking any particular coordinate system, and has a well-defined classical
meaning -- in fact, we will construct it first on smooth Riemannian spaces. It also scales, in the sense that the quantity we will 
define operationally is the ``quantum Ricci curvature at a given length scale". The seed of the idea owes much to the work
of Ollivier on discrete or coarse-grained Ricci curvature \cite{ollivier,ollivier1}. However, we had to alter his prescription 
quite substantially to make it suitable for application in nonperturbative quantum gravity. 

After giving a brief motivation for studying quantum implementations of curvature in the next section, 
we present our explicit construction of the classical version of quantum Ricci curvature in Sec.\ \ref{curv:sec}.
In Sec.\ \ref{smooth:sec} we perform a quantitative analysis of this quantity on smooth, two-dimensional,
constantly curved model spaces, in order to understand that the prescription is meaningful and to establish a
reference frame for the evaluation of the quantum Ricci curvature on non-smooth spaces.
Since our ultimate goal is the nonperturbative quantum theory, in the formulation of CDT, we then
implement our curvature construction on a variety of piecewise flat spaces. This allows us to understand the 
differences between continuum and discrete spaces and to quantify short-distance lattice artefacts.
Several regular lattices in two and three dimensions are treated in Sec.\ \ref{regular:sec}, 
and a variety of two-dimensional, equilateral 
random triangulations based on Delaunay triangulations in Sec.\ \ref{random:sec}. 
This demonstrates the computational feasibility of quantum Ricci curvature and illustrates its
behaviour under spatial averaging. Our conclusions and outlook are presented in Sec.\ \ref{conclusion:sec}.

\section{The case for (quantum) curvature}
\label{cqc:sec}

The key notion by which we understand and quantify the nontrivial local properties of classical spacetime is that of {\it curvature}. 
While most of our intuition about curvature comes from studying two-dimensional surfaces imbedded in three-di\-men\-sional
Euclidean space, intrinsic curvature in four dimensions -- as captured by the Riemann curvature tensor $R^\kappa{}_{\lambda\mu\nu}(x)$ --
is a complex and rather unintuitive quantity. Moreover, the components of the curvature tensor are not elementary, but derived quantities,
depending on the second derivatives of the metric tensor, which requires $g_{\mu\nu}(x)$ to be at least twice differentiable.
In situations where the metric structure is not of this type, it is clear that the definition of curvature will in general have to be modified 
to remain a meaningful concept. This will also be the case for the type of ``quantum geometry" we are interested in, which is
obtained as a continuum limit of an ensemble of piecewise flat geometries. 

We will introduce below a specific notion of
coarse-grained Ricci curvature that can be used in the context of nonperturbative, background-independent
quantum gravity. Like the dynamical dimensions mentioned above, the construction is in terms of measurements of volumes
and distances. As a consequence, it is robust and scalable, and can be implemented in a straightforward way on piecewise flat 
spaces, like those of Dynamical Triangulations. 

Note that we are not interested in investigating a classical limit in which a sequence of triangulated spaces approaches a given
smooth, classical metric manifold, and where one can then ask whether and how a particular notion of piecewise flat curvature 
converges to its smooth counterpart. 
Instead, in the gravitational path integral one considers a whole ensemble of 
spacetimes, and looks for continuum limits in which relevant observables exhibit a physically interesting scaling behaviour, 
and where most of the details of the regularization become irrelevant. 

There are already notions of curvature that have been used in this context, based on the concept
of a deficit angle, a simplicial implementation of describing the sectional curvature of two-dimensional subspaces. 
Regge calculus works with a standard expression for the scalar curvature in terms of deficit angles \cite{regge}.
Its integrated version appears in a simplicial analogue of the Einstein-Hilbert action, which is also used in quantum
Regge calculus and Dynamical Triangulations \cite{LRR}. 
In the context of Regge calculus, related simplicial representations have been constructed for more complicated 
curvature tensors (see, for example, \cite{hw,brewin,reggericci}). 

However, these expressions are not well suited as quantum observables in our sense.
The main issue is that the underlying notion of curvature defect is defined at the cutoff scale only. It does not scale
since there is no obvious way of defining a coarse-grained deficit angle. 
In nonperturbative quantum gravity models of the kind we are considering, integrated versions of the simplicial 
scalar curvature for $D>2$ tend to be highly divergent in the continuum limit. This happens
because the density of the curvature defects grows
as the lattice spacing $a$ goes to zero, while the individual deficit angles do not ``average out" on
coarse-grained scales. The alternative curvature observable we will define and implement in this
work both scales and has a better averaging behaviour, as we will demonstrate explicitly. 

The generalized notion of Ricci curvature introduced in the next section is not based on measuring deficit angles, 
but -- in a $D$-dimensional space -- involves the average distance between two overlapping $(D-1)$-dimensional spheres. 
The construction is inspired by the observation that on a smooth, compact Riemannian space with positive Ricci curvature,
the distance between two small spheres of radius $\epsilon$ is smaller than the distance between their two centres (see  
\cite{ollivier1} and references therein). The construction is entirely geometric and can be implemented in a straightforward
way on simplicial manifolds, using geodesic link distance (or dual geodesic link distance) and the natural volume measure, counting
the $D$-simplices. 
An important criterion that guided our search for a curvature observable is ease of implementation and low computational cost, 
in view of the fact that we are interested in evaluating it on non-infinitesimal neighbourhoods and 
in a quantum-gravitational context. Note that since it is natural to measure lengths and volumes in DT 
in terms of discrete units, measuring them is often reduced to counting, further simplifying implementation.

\begin{figure}[t]
\centerline{\scalebox{0.5}{\rotatebox{0}{\includegraphics{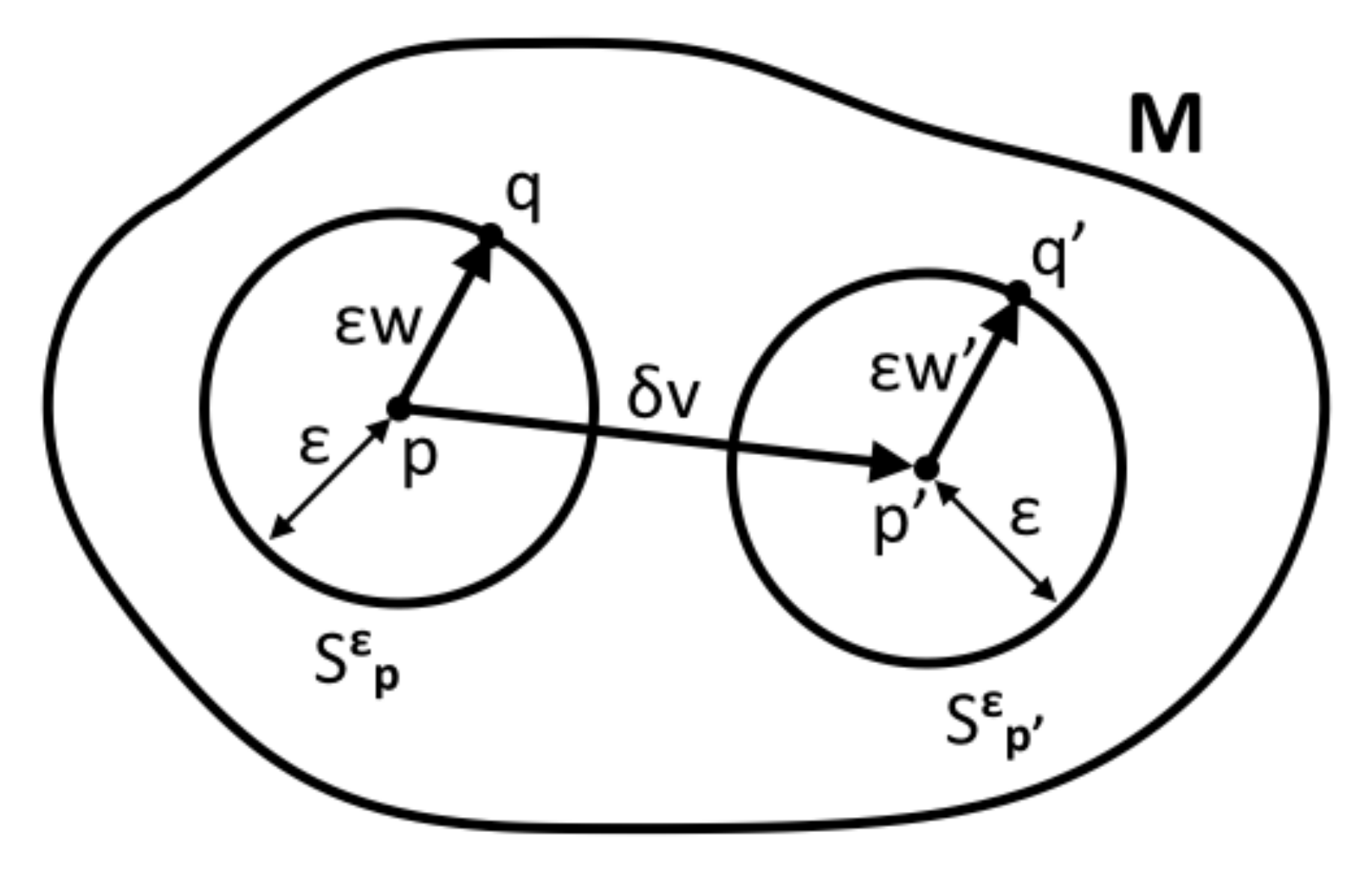}}}}
\caption{Two nearby spheres $S^\epsilon_p$ and $S^\epsilon_{p'}$ of radius $\epsilon$ whose centres are a small 
distance $\delta$ along the unit vector $v$ apart.
Parallel transport of a unit vector $w$ at $p$ along the geodesic of length $\delta$ connecting $p$ and $p'$ yields another unit vector $w'$.
The distance between the points $q$ and $q'$ in flat space is equal to $\delta$, while in the presence of curvature 
the lowest-order deviation from $\delta$ is given by formula (\ref{sectional}).}
\label{fig:2spheres}
\end{figure}

\section{A measure of curvature}
\label{curv:sec}

To motivate our construction, we begin with the case of a smooth, $D$-dimen\-sio\-nal Riemannian manifold $(M,g_{\mu\nu})$.
The eventual application we have in mind is DT or CDT quantum gravity (the latter after ``Wick rotation", such that 
the spacetimes summed over in
the path integral have positive definite metric \cite{physrep}). However, we do not see any obstacles to implementing
it in other discrete metric settings.
Consider two points $p$, $p'\in M$, which are a geodesic distance $\delta \geq 0$
apart. The two $\epsilon$-spheres $S_p^{\epsilon}$ and $S_{p'}^{\epsilon}$ around $p$ and $p'$ consist of those points in $M$ that are
a distance $\epsilon \geq 0$ away from the centres $p$ and $p'$ respectively. The parameters $\delta$ and $\epsilon$ must be
sufficiently small for the $\epsilon$-spheres to have the topology of $S^{D-1}$-spheres and such that $p'$ can be written
uniquely as $p'=\exp_p(\delta v)$ in terms of the exponential map, where $v$ is a unit vector in the tangent space $T_pM$.

There are various ways of defining the distance between the two spheres $S^\epsilon_p$ and $S^\epsilon_{p'}$.
For example, parallel transport of tangent vectors in $T_pM$ to $T_{p'}M$ along the geodesic connecting the centres $p$ and $p'$ can
be used to identify points on the two spheres pairwise in a unique way. Suppose $q$ is the image $q=\exp_p(\epsilon w)$
of the point $p$ under the exponential mapping in the direction of the unit vector $w\in T_pM$. Parallel-transporting the vector $w$ to $p'$
yields another unit vector, $w'\in T_{p'}M$, 
which therefore points to a point $q'$ on the sphere $S_{p'}^\epsilon$ in the sense that $q'=\exp_{p'}(\epsilon w')$
(see Fig.\ \ref{fig:2spheres}).
Again, for this construction to be well-defined and unique, both $\delta$ and $\epsilon$ must be sufficiently small, which we will
assume is the case. Assuming for simplicity that $v$ and $w$ are orthogonal, the distance between the two points $q$ and $q'$ 
is given by \cite{ollivier}
\begin{equation}
d(q,q')=\delta \bigg(1-\frac{\epsilon^2}{2}K(v,w)+O(\epsilon^3+\delta \epsilon^2)\bigg),
\label{sectional}
\end{equation}  
in the limit $(\delta,\epsilon)\rightarrow (0,0)$, where the sectional curvature $K(v,w)$ is the Gaussian curvature of the two-dimensional 
surface of geodesics whose tangent vector at $p$ lies in the span of $v$ and $w$.
Recall that the sectional curvature is defined in terms of the Riemann curvature tensor $R$ and the scalar product 
$\langle\cdot,\cdot \rangle$ on $M$ as
\begin{equation}
K(v,w)=\frac{ \langle R(v,w)w,v\rangle}{\langle v,v\rangle \langle w,w \rangle-\langle v,w\rangle^2},
\label{secdef}
\end{equation}
where the denominator simplifies to 1 for orthonormal vectors $v$ and $w$.

If $S_p^{\epsilon}$ is mapped to $S_{p'}^{\epsilon}$ using parallel transport, as described above,
a point $q\in S_p^{\epsilon}$ and its image $q'\in S_{p'}^{\epsilon} $ are on average a distance 
\begin{equation}
d(S_p^{\epsilon},S_{p'}^{\epsilon})
=  \delta \bigg(1-\frac{\epsilon^2}{2D}\, Ric(v,v)+O(\epsilon^3+\delta \epsilon^2)\bigg),
\label{ricci}
\end{equation}  
apart in the limit $(\delta,\epsilon)\rightarrow (0,0)$ \cite{ollivier1}. On the right-hand side of (\ref{ricci}), $Ric(v,v)$ denotes the
Ricci curvature associated with the unit vector $v$, given as the sum of the sectional curvatures of all planes containing $v$.  
In terms of an orthonormal basis $\{ e_i, \, i=1,\dots,D \}$, it can be written as 
\begin{equation}
Ric(e_1,e_1)=\sum_{i=2}^D K(e_1,e_i),
\label{ksum}
\end{equation}
say. Expression (\ref{ricci}) is obtained by integrating the point distances corresponding to all unit vectors $w$ and dividing by
the volume of the unit $(D-1)$-sphere. Equivalently, one can integrate over the sphere of radius $\epsilon$ with respect to
the induced measure and divide by the sphere volume.

For a variety of reasons the prescription (\ref{ricci}) is not particularly suited to extracting curvature from a simplicial manifold. 
Although the underlying parallel transport can be implemented straightforwardly, as was shown in \cite{wilson} for the case of dynamical 
triangulations, local coordinate systems generally cannot be extended beyond pairs of adjacent building blocks. Geodesics
between vertices further than one unit distance apart are in general non-unique. Moreover, if we consider only the distances between 
the {\it vertices} contained in nearby $\epsilon$-spheres -- as is natural in a simplicial context -- their number will typically be 
different for different spheres.\footnote{We should put ``spheres" in inverted commas here, since the vertices and other (sub-) simplices at
integer link distance $\epsilon$ from a chosen vertex do not in general form a topological $(D-1)$-sphere, but a disconnected space.}
This means that we cannot associate them pairwise in a one-to-one fashion, as was done to obtain (\ref{ricci}).

Instead of the sphere distance (\ref{ricci}) on smooth manifolds, we will use a more robust notion of ``average sphere distance" that can 
be implemented also on more general metric spaces and not just in the limit of small distances. For this purpose, we will from now on interpret an
``$\epsilon$-sphere" $S^\epsilon_p$ as the set of all points at distance $\epsilon$ from a given centre point $p$, regardless of
whether they form a sphere topologically. On a $D$-dimensional Riemannian manifold, the average sphere distance $\bar d$
of two $\epsilon$-spheres centred at points $p$ and $p'$ is simply defined as
\begin{equation}
\bar{d}(S_p^{\epsilon},S_{p'}^{\epsilon}):=\frac{1}{vol(S_p^{\epsilon})}\frac{1}{vol(S_{p'}^{\epsilon})}
\int_{S_p^{\epsilon}}d^{D-1}q\; \sqrt{h} \int_{S_{p'}^{\epsilon}}d^{D-1}q'\; \sqrt{h'}\ d(q,q'),
\label{sdist}
\end{equation}   
where $h$ and $h'$ are the determinants of the metrics induced on the two $(D-1)$-dimensional ``spheres", which are also used to compute the
sphere volumes $vol(S)$, and $d(q,q')$ denotes the geodesic distance between the points $q$ and $q'$. Note that $\bar d$ is not a proper
distance in the mathematical sense. Although it is positive, symmetric and satisfies the triangle inequality, the average distance of an 
$\epsilon$-sphere to itself does not vanish, unless $\epsilon =0$. 

For the definition (\ref{sdist}) to be meaningful, it is not essential 
that the two spheres have the same radius, but this is the only case we will consider in the following. More specifically, 
our definition of ``quantum Ricci curvature"
will be based on pairs of overlapping spheres whose common radius is equal to the distance between their centres, $\epsilon =\delta$. 
This is not a unique
choice, but a natural and convenient one if one is interested in a scalable curvature observable associated with a single scale $\delta$.
Adapting the definition (\ref{sdist}) to a piecewise flat manifold made from equilateral simplices (a typical DT configuration), we
have
\begin{equation}
\bar{d}(S_p^{\epsilon},S_{p'}^{\epsilon})=\frac{1}{N_0(S_p^{\epsilon})}\frac{1}{N_0(S_{p'}^{\epsilon})}
\sum_{q\in S_p^{\epsilon}} \sum_{q'\in S_{p'}^{\epsilon}} d(q,q'),
\label{simpdist}
\end{equation}   
where $N_0(S^\epsilon_p)$ is the number of vertices at link distance $\epsilon$ from the central vertex $p$ and $d(q,q')$ now denotes
the geodesic link distance between the vertices $q$ and $q'$, i.e. the number of links in the shortest path along links from $q$ to $q'$. 

We will extract a notion of {\it quantum Ricci curvature}\footnote{The explicit construction and implementations 
described in what follows are not primarily of a quantum nature. However, the motivation and intended main application 
of this work is nonperturbative quantum gravity, justifying the label ``quantum" (see also Sec.\ \ref{conclusion:sec} for
further comments). A genuine quantum-gravitational application will be presented in \cite{qrc2}.} $K_q(p,p')$, 
associated with a pair of points $p$ and $p'$ separated by a 
distance $\delta$, 
by comparing the average distance of the two $\delta$-spheres centred at $p$ and $p'$ with $\delta$ according to
\begin{equation}
\frac{\bar{d}(S_p^{\delta},S_{p'}^{\delta})}{\delta}=c_q (1 - K_q(p,p')),\;\;\; \delta =d(p,p'),
\label{qric}
\end{equation}
where $c_q$ is a positive constant, which depends on the metric space under consideration,
and $K_q$ captures any nontrivial dependence on $\delta$. 
This construction is similar to Ollivier's ``coarse Ricci curvature" \cite{ollivier}, specialized to a pair of spheres, but
using the average distance (\ref{sdist}) or (\ref{simpdist}) instead of the $L^1$-transportation distance.
The latter is a genuine distance, but expensive to compute, because it is defined through an optimization. 
Since computability is an important requirement for the application we have in mind, we are using the
average distance instead. In the next section, we will evaluate $\bar{d}/\delta$ for some classical, constantly curved model spaces and
for non-infinitesimal distances and
show that -- unlike for the corresponding quantity in \cite{ollivier} -- the constant $c_q$ in (\ref{qric}) in general is
not equal to 1.

\section{Smooth model spaces}
\label{smooth:sec}

To develop a better understanding for the generalized Ricci curvature $K_q$, we will begin by evaluating it on smooth,
constantly curved Riemannian manifolds, starting with the flat, spherical and hyperbolic spaces in $D=2$ dimensions.

\begin{figure}[t]
\begin{tabular}{ll}
\includegraphics[scale=0.65]{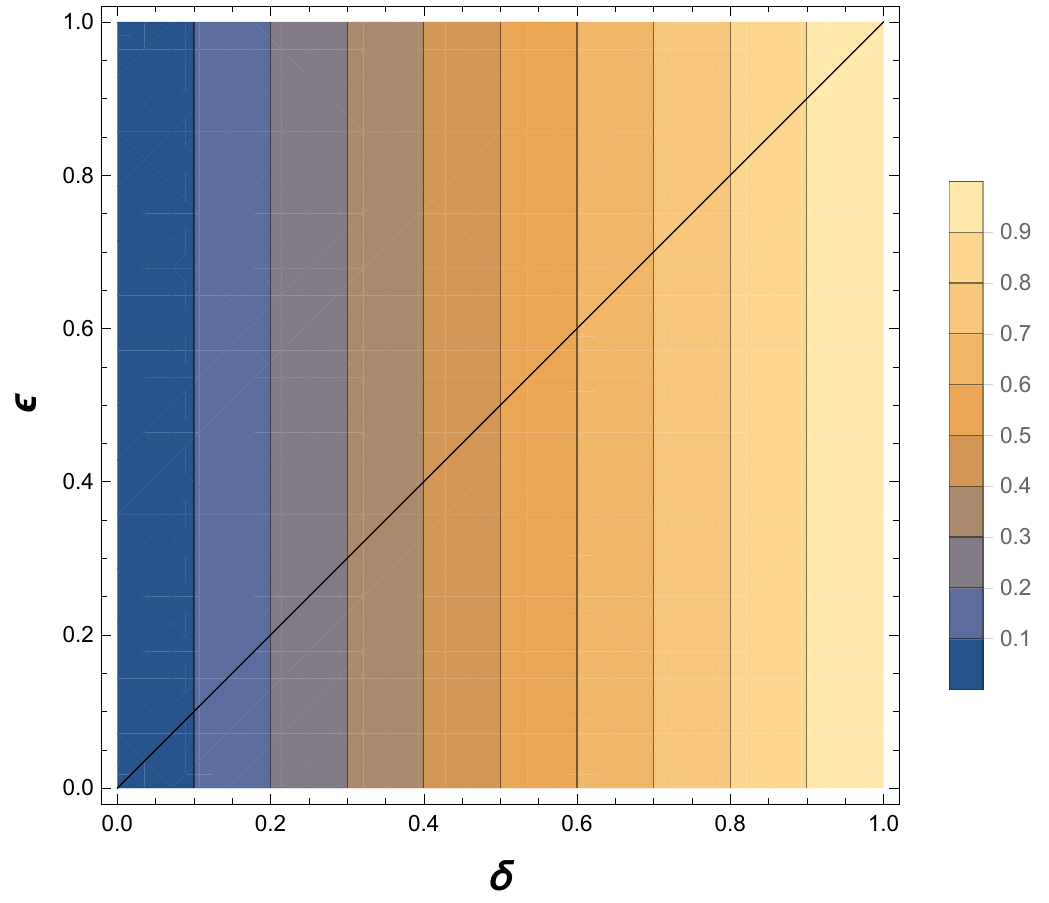}
&
\includegraphics[scale=0.65]{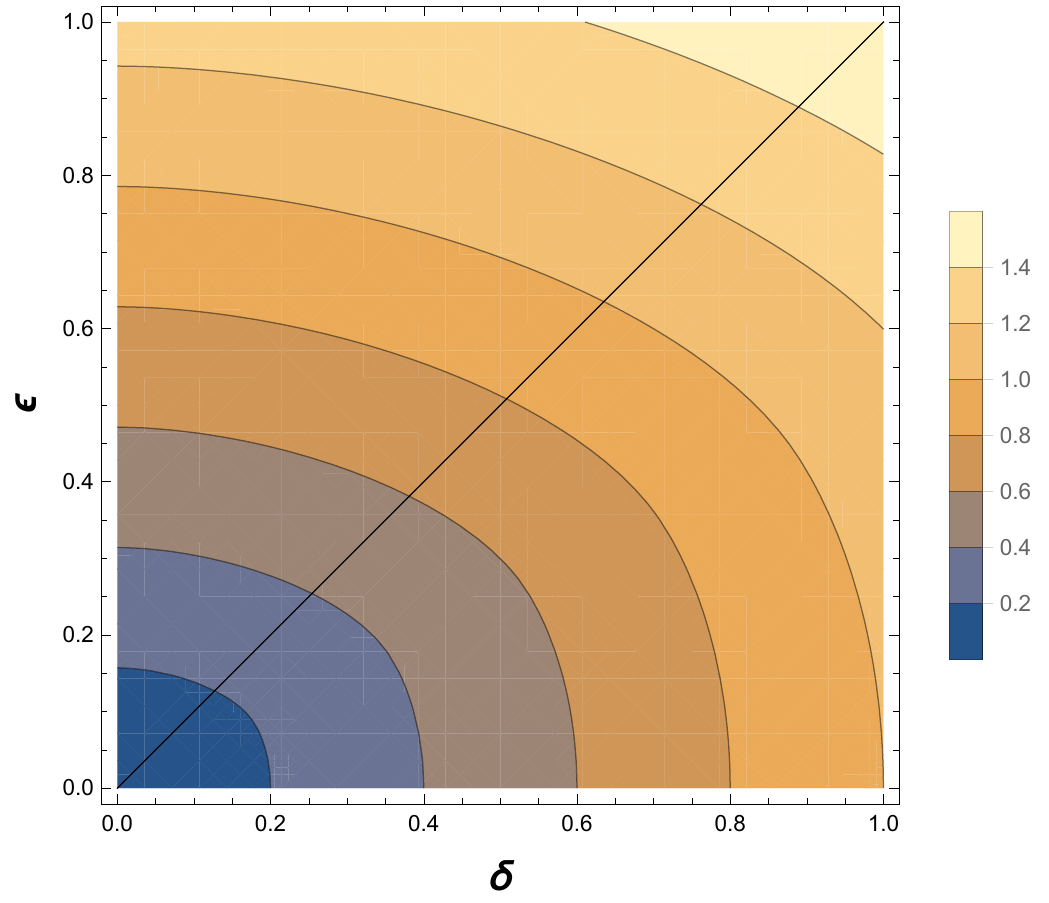}
\end{tabular}
\caption{Contour plots of the distance between two circles on a two-dimensional
flat space, as function of the circle radius $\epsilon$ and
the distance $\delta$ of their centres. Left: sphere distance (\ref{flat2d}). 
Right: average sphere distance (\ref{2dflatav}).}
\label{fig:flat}
\end{figure}

Consider a pair of spheres (circles) of radius $\epsilon$ in two-dimensional flat, Euclidean space, whose centres are a
distance $\delta$ apart, not necessarily equal to $\epsilon$. We will use an angular parameter $\alpha\in [0,2\pi [$ to 
uniquely label points $q_\alpha$ along the sphere $S_p^\epsilon$ and the corresponding points $q_\alpha'$ on
$S_{p'}^\epsilon$. Our convention is to measure the angle around $p$ in anticlockwise direction from the geodesic
connecting $p$ and $p'$.
Otherwise the situation is like that depicted in Fig.\ \ref{fig:2spheres}. For the sphere distance
we compute
\begin{equation}
d(S_p^{\epsilon},S_{p'}^{\epsilon})=\frac{1}{2\pi} \int_0^{2\pi}\!\!\! d\alpha\; d(q_\alpha,q_\alpha')=
\frac{1}{2\pi} \int_0^{2\pi} \!\!\! d\alpha\; \delta =\delta,
\label{flat2d}
\end{equation} 
independent of $\epsilon$, since in flat space all point pairs $(q_\alpha,q_\alpha')$ are exactly a distance $\delta$ apart. 
The result (\ref{flat2d}) is consistent with the right-hand side of eq.\ (\ref{ricci}), because in flat space $Ric(v,v)$ vanishes 
identically for all vectors $v$. The computation of the average distance of the two spheres involves a double integral,
\begin{eqnarray} 
{\hspace{-0.2cm}}\!\!\!\!\!
\bar{d}(S_p^{\epsilon},S_{p'}^{\epsilon})\!\!\!\! & =&\! \!\!\! \frac{1}{(2\pi)^2} \int_0^{2\pi}\!\!\! d\alpha \int_0^{2\pi}\!\!\! d\beta\; d(q_\alpha,q_\beta') \nonumber \\
&=&\!\!\!\! \frac{1}{(2\pi)^2} \int_0^{2\pi}\!\!\! d\alpha \int_0^{2\pi}\!\!\! d\beta 
\sqrt{(\delta+\epsilon (\cos \beta -\cos \alpha))^2+\epsilon^2 (\sin\beta-\sin\alpha)^2},
\label{2dflatav}
\end{eqnarray}
where the two angles $\alpha$ and $\beta$ label arbitrary 
pairs of points $(q_\alpha,q_\beta')\in S_p^\epsilon\times S_{p'}^\epsilon$.
Since we are unable to evaluate the integrals in (\ref{2dflatav}) analytically, we resort to a numerical evaluation. 
Fig.\ \ref{fig:flat} shows contour plots of the sphere distance and the average sphere distance as functions 
of $\delta$ and $\epsilon$. Note that for $\delta =\epsilon$, the case we will
be considering for the quantum Ricci curvature, expression (\ref{2dflatav}) is purely linear in $\delta$. 
This corresponds to the diagonal in the plot on the right in Fig.\ \ref{fig:flat}. Numerically, the average sphere distance
in this case evaluates to 
\begin{equation}
\bar{d}(S_p^{\delta},S_{p'}^{\delta}) \approx 1.5746\, \delta\;\;\;\; {\rm (flat\; case).}
\label{dflat}
\end{equation}
Comparing with the sphere distance of relation (\ref{flat2d}), we see that the constant prefactor of $\delta$ has changed from
1 to $c_q\approx 1.5746$. For smooth Riemannian mani\-folds, the
coefficient of $\delta$ in the power expansion of $\bar{d}(S_p^{\delta},S_{p'}^{\delta})$ is universal and depends only on 
the dimension of $M$. For instance, an analogous calculation for the average sphere distance (for $\epsilon\! =\!\delta$) 
in three-dimensional flat space yields $\bar{d}\approx 1.6250\, \delta$.

Next, we consider pairs of $\epsilon$-circles on a constantly curved two-sphere of embedding radius $\rho$. In other words,
we can think of the two-sphere as consisting of all points $(x,y,z)\in\R^3$ satisfying $x^2+y^2+z^2=\rho^2$. Points on the two-sphere
can also be described by two angles $\theta\in [0,\pi]$ and $\varphi \in [0,2\pi [$. Recall that geodesics on $S^2$ are arcs of great circles
and that the geodesic distance between two points $(\theta_i,\varphi_i)$, $i=1,2$, is given by
\begin{equation}
d((\theta_1,\varphi_1),(\theta_2,\varphi_2))=\rho\,\arccos (\cos\theta_1\cos\theta_2+\sin\theta_1\sin\theta_2\cos(\varphi_2-\varphi_1)).
\label{sphdist}
\end{equation}
The sphere distance of two $\epsilon$-circles whose centres are a distance $\delta$ apart is given by
\begin{equation} 
d(S_p^{\epsilon},S_{p'}^{\epsilon})=\frac{1}{2\pi} \int_0^{2\pi}\!\!\! d\alpha\;\rho\,\arccos \Big(\cos\tfrac{\delta}{\rho}+\sin^2\!\alpha\,
\sin^2 (\tfrac{\epsilon}{\rho})\, (1-\cos\tfrac{\delta}{\rho})\Big).
\label{round2d}
\end{equation}
Expanding the integrand on the right-hand side of (\ref{round2d}), which is the curved-space analogue of the
distance $d(q_\alpha,q_\alpha')$ in the flat-space integral (\ref{flat2d}),
for small $\delta$ and $\epsilon$ leads to
 \begin{eqnarray}
&& {\hspace{-0.2cm}}\!\!\!\!\!\!
 \frac{\delta}{2\pi}\int_0^{2\pi}\!\!\! d\alpha \left( 
  1-\!\tfrac{1}{2}\big(\tfrac{\epsilon}{\rho}\big)^2 \sin^2\! \alpha+(\tfrac{\epsilon}{\rho})^4\left(\tfrac{1}{6}\, 
  \sin^2\! \alpha -\!\tfrac{1}{8}\, \sin^4\! \alpha\right)-\!
  \tfrac{1}{24}\big(\tfrac{\epsilon}{\rho}\big)^2\big(\tfrac{\delta}{\rho}\big)^2 \sin^2\!\alpha +\! {\rm h.o.}\right)\nonumber \\
  &&\;\;\;\;\;\;\;\; =\delta \left(1-\tfrac{1}{4}\big( \tfrac{\epsilon}{\rho}\big)^2+
   \tfrac{7}{192}\big(\tfrac{\epsilon}{\rho}\big)^4-
   \tfrac{1}{48}\big(\tfrac{\epsilon}{\rho}\big)^2\big(\tfrac{\delta}{\rho}\big)^2+ {\rm h.o.}\right)\! ,
\label{sphexp}   
\end{eqnarray}
where h.o. denotes terms of combined $\delta$- and $\epsilon$-order of at least six. Given that the Ricci curvature
$Ric(v,v)$ on the two-sphere is $1/\rho^2$ for any unit vector $v$, we see that the integration result in (\ref{sphexp})
is consistent with the general formula (\ref{ricci}) for $D=2$.
\begin{figure}[t]
\begin{tabular}{ll}
\includegraphics[scale=0.65]{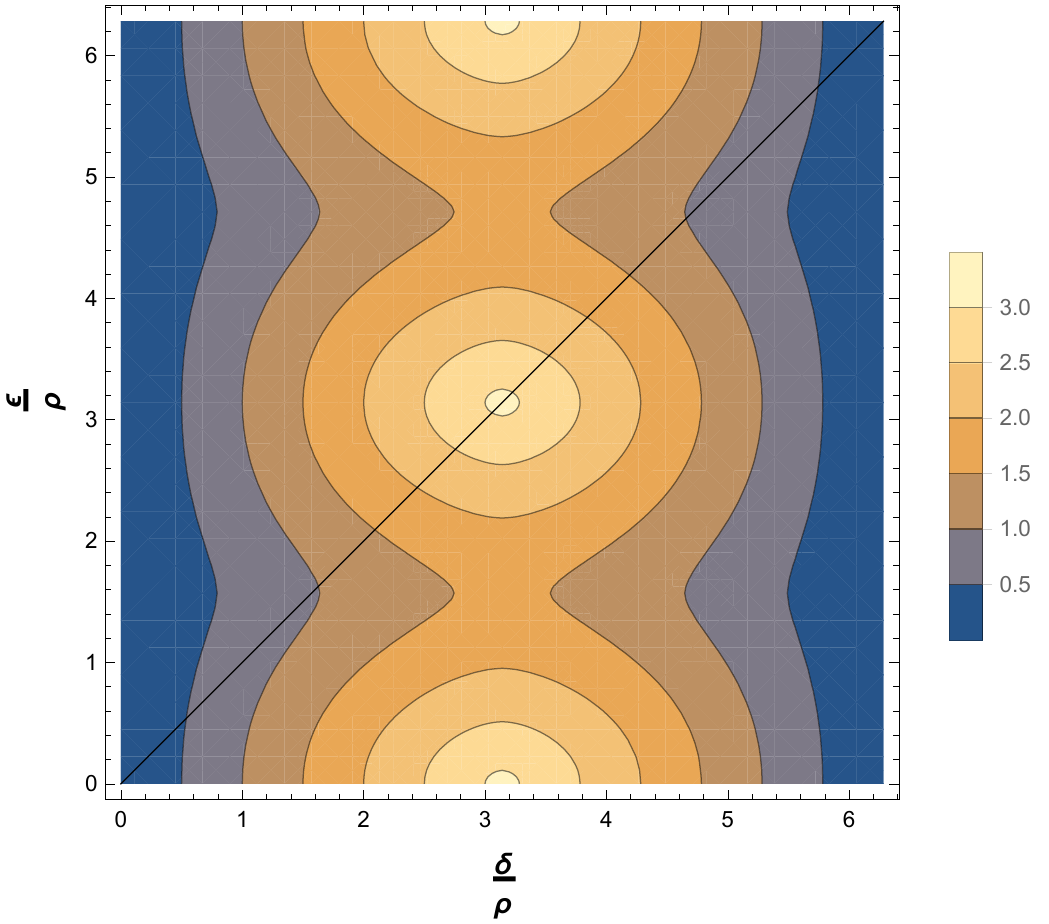}
&
\includegraphics[scale=0.65]{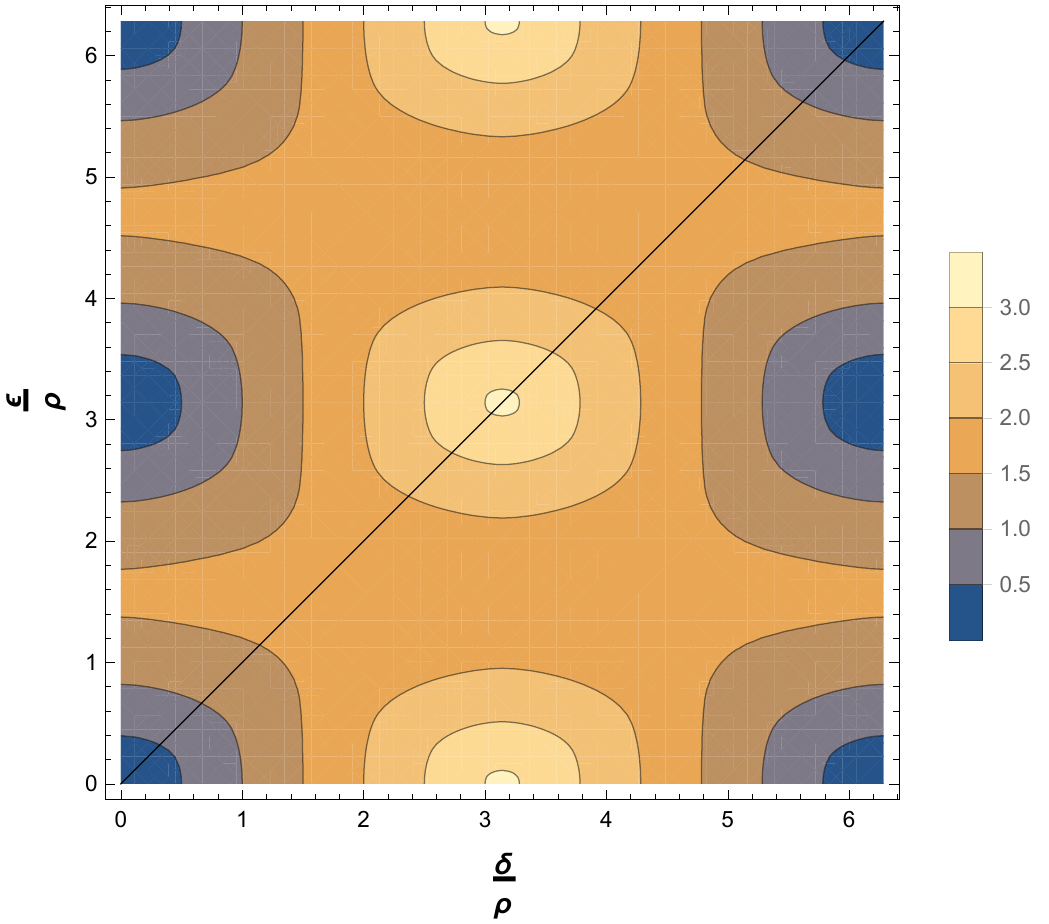}
\end{tabular}
\caption{Contour plots of the distance between two circles on a two-dimensional
space of constant positive curvature, as function of the circle radius $\epsilon$ and
the distance $\delta$ of their centres, both rescaled by the curvature radius $\rho$. Left: sphere distance (\ref{round2d}). 
Right: average sphere distance (\ref{round2dav}).}
\label{fig:sphere}
\end{figure}
Considering next the average sphere distance and again using eq.\ (\ref{sphdist}), we find
\begin{eqnarray}
{\hspace{-0.2cm}}\!\!\!\!\!
&&\!\!\!\!\!\!\!\!\!  \bar{d}(S_p^{\epsilon},S_{p'}^{\epsilon}) = 
 \frac{\rho}{(2\pi)^2}\int_{0}^{2\pi}\!\!\! d\alpha\int_{0}^{2\pi}\!\!\!d\beta\, \arccos\bigg(
 \sqrt{1-\sin^2(\tfrac{\epsilon}{\rho})\sin^2\!\alpha}\sqrt{1-\sin^2(\tfrac{\epsilon}{\rho})\sin^2\!\beta}\nonumber\\
  &&\!\!\!\!\!\!\!\! \times \cos\big(\tfrac{\delta}{\rho}+\!\arctan\big(\!\tan\tfrac{\epsilon}{\rho}\cos\beta \big)\! -\!
  \arctan\big(\!\tan\tfrac{\epsilon}{\rho}\cos\alpha \big)\big)\! + \!
  \sin^2(\tfrac{\epsilon}{\rho})\sin\alpha\,\sin\beta   \bigg)\! .
\label{round2dav}  
 \end{eqnarray}
Setting $\epsilon =\delta$ and expanding this expression for small $\delta$ results in
\begin{equation}
\bar{d}(S_p^{\delta},S_{p'}^{\delta}) \approx \delta\Big(1.5746-0.1440 \big(\tfrac{\delta}{\rho}\big)^2
-0.0239 \big(\tfrac{ \delta}{\rho} \big)^4+O\Big(\big(\tfrac{\delta}{\rho}\big)^6\Big)\Big),
\label{sphdeltaexp}
\end{equation}
where the coefficients were determined by numerical integration. At linear order in $\delta$ the same constant appears
as in the flat case of eq.\ (\ref{dflat}). The next-to-leading order is proportional to the Ricci curvature with a negative
coefficient, which is qualitatively similar to the behaviour (\ref{sphexp}) of the corresponding expansion of the sphere distance. 
The contour plots for both types of sphere distance are shown in Fig.\ \ref{fig:sphere}, as functions of $\delta/\rho$ and
$\epsilon/\rho$, both taking values in the interval $[0,2\pi]$. Note that the plots are symmetric under both
$\delta\mapsto 2\pi\rho-\delta$ and $\epsilon\mapsto 2\pi\rho-\epsilon$, a property that can be read off easily from their analytic
expressions (\ref{round2d}) and (\ref{round2dav}).

\begin{figure}[t]
\begin{tabular}{ll}
\includegraphics[scale=0.65]{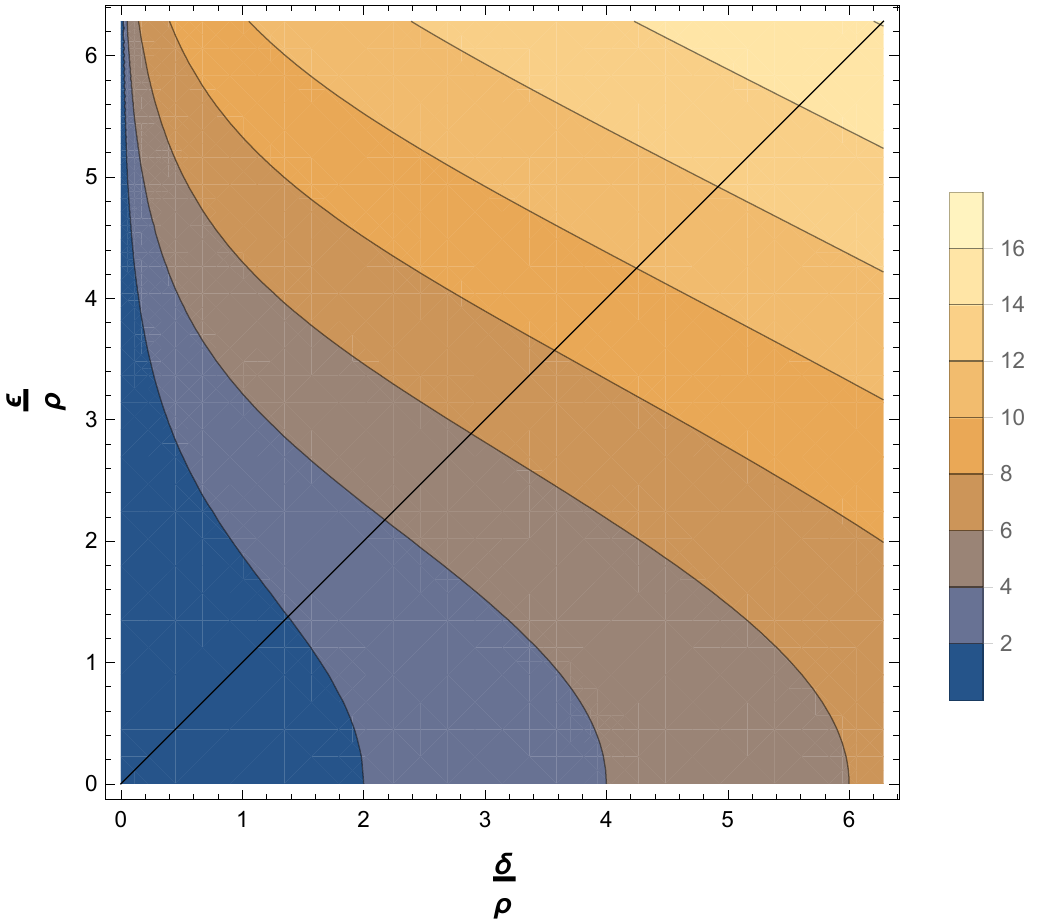}
&
\includegraphics[scale=0.65]{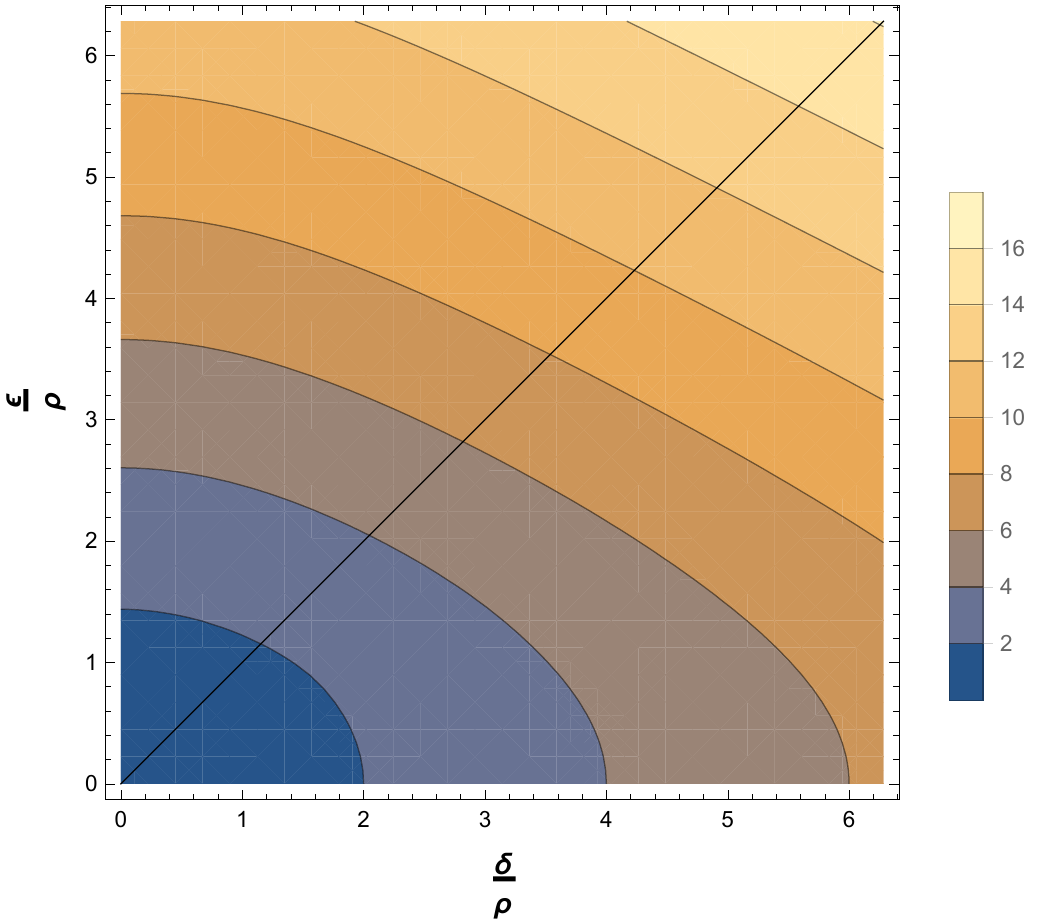}
\end{tabular}
\caption{Contour plots of the distance between two circles on a two-dimensional
space of constant negative curvature, as function of the circle radius $\epsilon$ and
the distance $\delta$ of their centres, both rescaled by the curvature radius $\rho$. Left: sphere distance (\ref{hyper2d}). 
Right: average sphere distance (\ref{hyp2dav}).}
\label{fig:hyper}
\end{figure}
Lastly, we consider circle distances on the negatively curved hyperbolic space in two dimensions, defined as the set of all points
$(x,y,z)\in\R^3$ satisfying $-x^2-y^2+z^2=\rho^2$ and $z>0$.
Points on this space can be parametrized by a hyperbolic angle $\theta\in [0,\infty [$ and an ordinary angle $\varphi\in[0,2\pi [$,
which are related to the Euclidean coordinates by $\theta ={\rm arccosh} (z/\rho)$ and $\varphi=\arctan (y/x)$.
The geodesic distance between two such points $(\theta_i,\varphi_i)$, $i=1,2$, is given by
\begin{equation}
d((\theta_1,\varphi_1),(\theta_2,\varphi_2))=\rho\,\rm{arccosh} (\cosh\theta_1\cosh\theta_2-\sinh\theta_1\sinh\theta_2\cos(\varphi_2-\varphi_1)).
\label{hypdist}
\end{equation}
From this, we obtain
the sphere distance of two $\epsilon$-circles at distance $\delta$ as a straightforward hyperbolic version of formula (\ref{round2d}),
namely,
\begin{equation} 
d(S_p^{\epsilon},S_{p'}^{\epsilon})=\frac{1}{2\pi} \int_0^{2\pi}\!\!\! d\alpha\;\rho\,{\rm arccosh} (\cosh\tfrac{\delta}{\rho}+\sin^2\!\alpha\,
\sinh^2 (\tfrac{\epsilon}{\rho})\, (\cosh\tfrac{\delta}{\rho}-1)),
\label{hyper2d}
\end{equation}
whose expansion for small $\delta$ and $\epsilon$ is given by
 \begin{eqnarray}
&& {\hspace{-0.2cm}}\!\!\!\!\!\!
 \frac{\delta}{2\pi}\int_0^{2\pi}\!\!\! d\alpha \left( 
  1+\!\tfrac{1}{2}\big(\tfrac{\epsilon}{\rho}\big)^2 \sin^2\! \alpha+(\tfrac{\epsilon}{\rho})^4\left(\tfrac{1}{6}\, 
  \sin^2\! \alpha -\!\tfrac{1}{8}\, \sin^4\! \alpha\right)-\!
  \tfrac{1}{24}\big(\tfrac{\epsilon}{\rho}\big)^2\big(\tfrac{\delta}{\rho}\big)^2 \sin^2\!\alpha +\! {\rm h.o.}\right)\nonumber \\
  &&\;\;\;\;\;\;\;\; =\delta \left(1+\tfrac{1}{4}\big( \tfrac{\epsilon}{\rho}\big)^2+
   \tfrac{7}{192}\big(\tfrac{\epsilon}{\rho}\big)^4-
   \tfrac{1}{48}\big(\tfrac{\epsilon}{\rho}\big)^2\big(\tfrac{\delta}{\rho}\big)^2+ {\rm h.o.}\right)\! ,
\label{hypexp}   
\end{eqnarray}
which to this order is identical with the corresponding formula (\ref{sphexp}), apart from the sign of the term proportional
to $\delta\epsilon^2$. Comparing this term with eq.\ (\ref{ricci}) leads to $Ric(v,v)=-1/\rho^2$, which 
is the well-known result for the Ricci curvature on hyperbolic space. The average sphere distance in hyperbolic space
is given by the double integral
\begin{eqnarray}
{\hspace{-0.2cm}}\!\!\!\!\!
&&\!\!\!\!\!\!\!\!\!  \bar{d}(S_p^{\epsilon},S_{p'}^{\epsilon}) = 
 \frac{\rho}{(2\pi)^2}\int_{0}^{2\pi}\!\!\! d\alpha\int_{0}^{2\pi}\!\!\!d\beta\, {\rm arccosh} \bigg(\! 
(\cos\beta-\cos\alpha) \sinh\tfrac{\epsilon}{\rho} \cosh\tfrac{\epsilon}{\rho} \sinh\tfrac{\delta}{\rho}  
\nonumber\\
  && \hspace{0.7cm}
    +\cosh^2(\tfrac{\epsilon}{\rho}) \cosh\tfrac{\delta}{\rho}    
  -\sinh^2(\tfrac{\epsilon}{\rho})\Big( \sin\alpha\, \sin\beta+\cos\alpha\,\cos\beta\, \cosh\tfrac{\delta}{\rho}   \Big)
     \bigg)\! .
\label{hyp2dav}  
 \end{eqnarray}
Setting $\epsilon =\delta$ and expanding this expression for small $\delta$ gives
\begin{equation}
\bar{d}(S_p^{\delta},S_{p'}^{\delta}) \approx \delta\Big(1.5746+0.1440 \big(\tfrac{\delta}{\rho}\big)^2
-0.0239 \big(\tfrac{ \delta}{\rho} \big)^4+O\Big(\big(\tfrac{\delta}{\rho}\big)^6\Big)\Big),
\label{hypdeltaexp}
\end{equation}
where the coefficients were again determined by numerical integration. Comparing this with the corresponding
expansion (\ref{sphdeltaexp}) for the spherical case, we observe the same behaviour as for the sphere distances:
the expansions to this order are the same, only the term proportional to $\delta^3$ has its sign reversed because of the opposite sign
of the Ricci curvature. 

\begin{figure}[t]
\begin{tabular}{ll}
\includegraphics[scale=0.53]{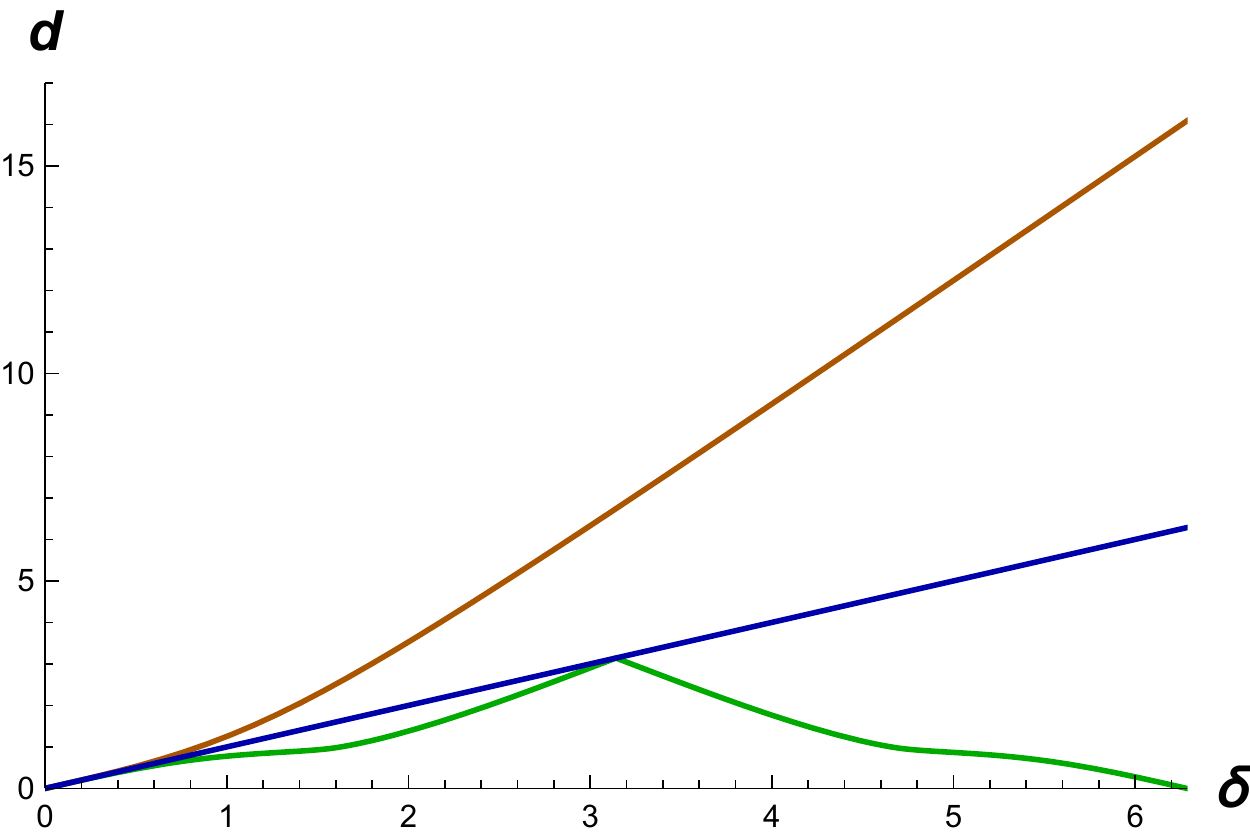}
&
\includegraphics[scale=0.53]{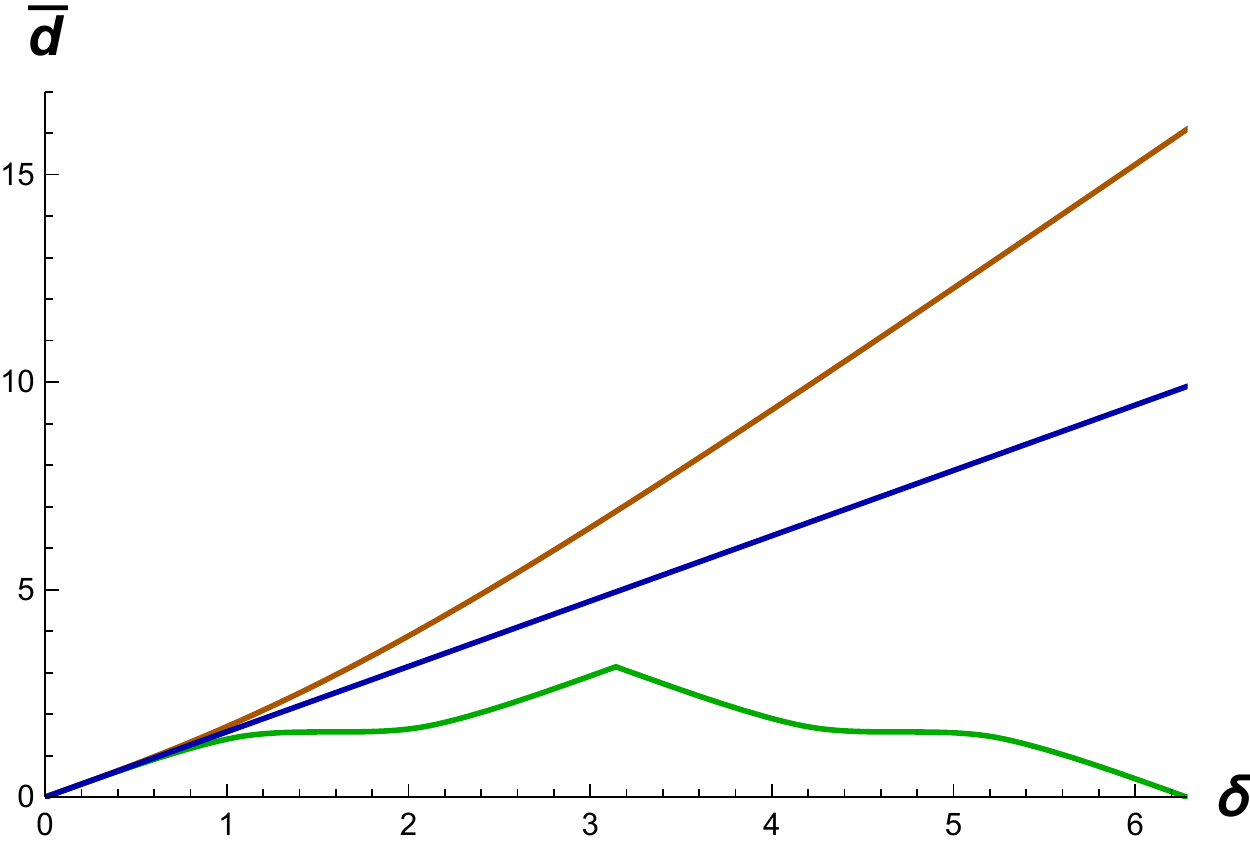}
\end{tabular}
\caption{Comparing sphere distance (left) and average sphere distance (right) for $\epsilon\! =\!\delta$, as function of $\delta\in[0,2\pi ]$,
for the three constantly
curved model spaces: hyperbolic (top), flat (middle) and spherical (bottom). The curvature radius $\rho$ has been set to 1.}
\label{fig:diag}
\end{figure}
\begin{figure}[t]
\begin{tabular}{ll}
\includegraphics[scale=0.53]{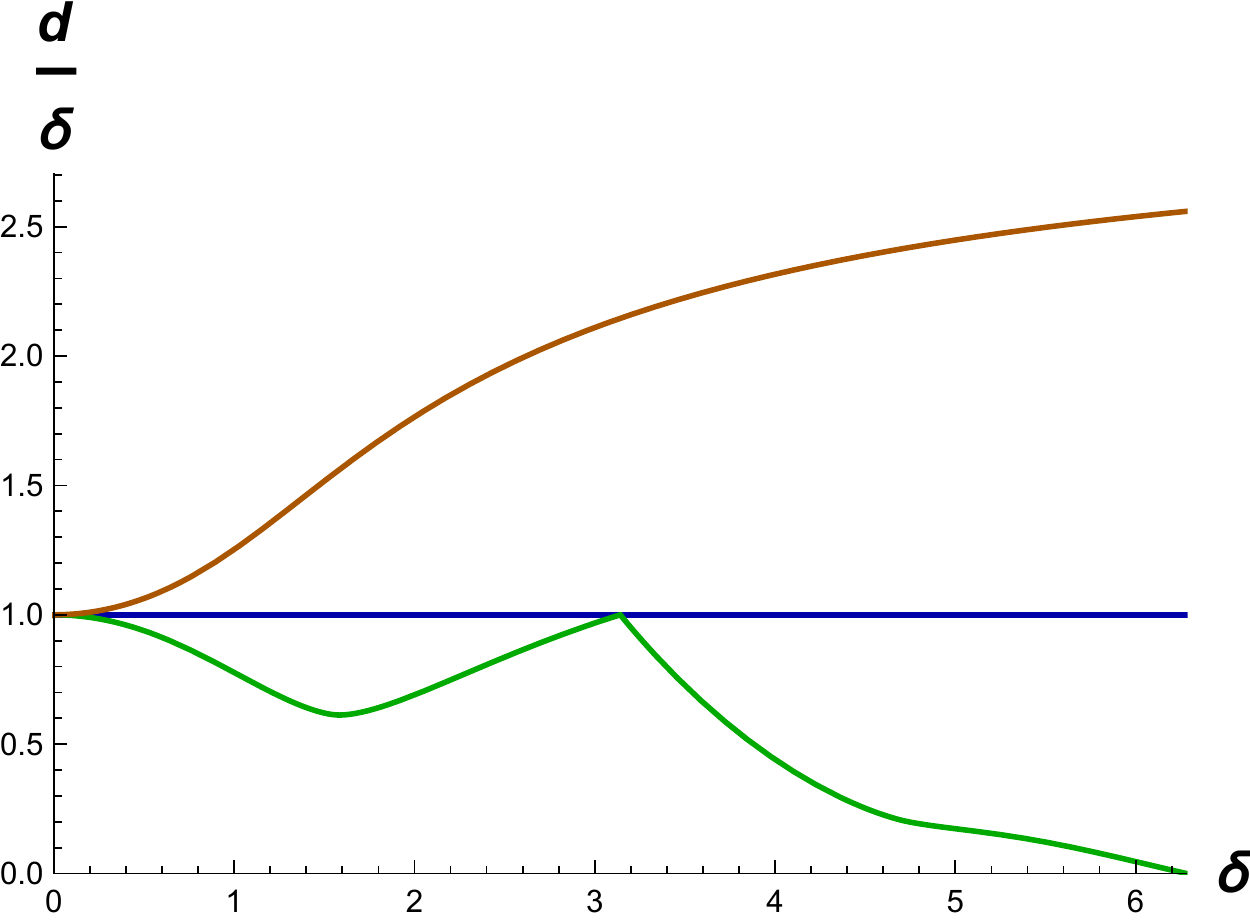}
&
\includegraphics[scale=0.53]{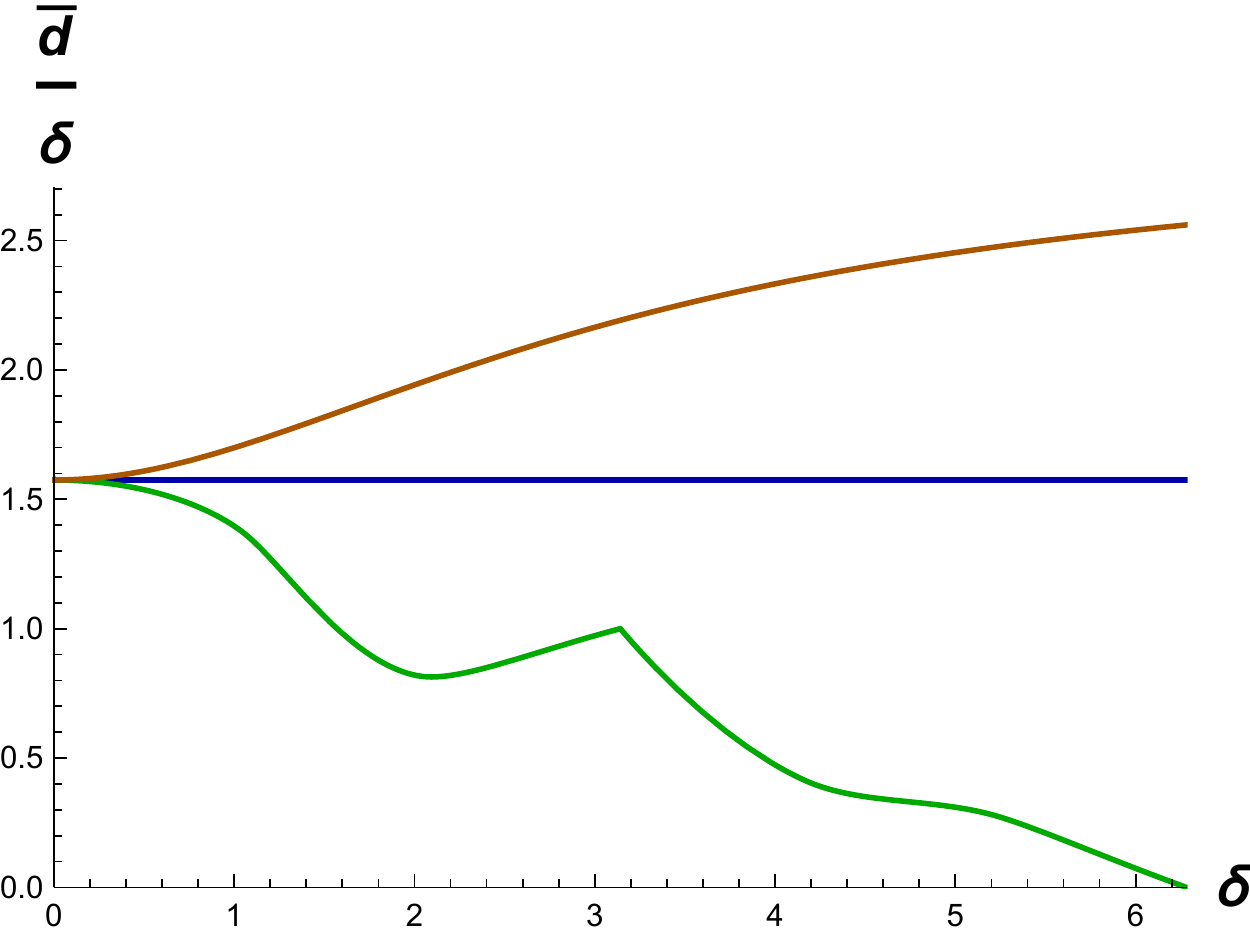}
\end{tabular}
\caption{Comparing the normalized versions of sphere distance (left) and average sphere distance (right) for $\epsilon\! =\!\delta$,
as function of $\delta\in[0,2\pi ]$, for the three constantly
curved model spaces: hyperbolic (top), flat (middle) and spherical (bottom). The curvature radius $\rho$ has been set to 1.}
\label{fig:diagnorm}
\end{figure}

\noindent Fig.\ \ref{fig:hyper} juxtaposes the behaviour of the sphere distance and the average sphere distance
as functions of both $\delta/\rho$ and $\epsilon/\rho$. The ranges of these hyperbolic angles have been chosen identical to the sphere
case of Fig.\ \ref{fig:sphere} for ease of comparison.  
The three plot pairs of Figs.\ \ref{fig:flat}, \ref{fig:sphere} and \ref{fig:hyper} illustrate the behaviour of the two
sphere distances $d(S_p^{\epsilon},S_{p'}^{\epsilon})$ and $\bar{d}(S_p^{\epsilon},S_{p'}^{\epsilon})$ for large values of 
$\delta$ and $\epsilon$, and specifically the qualitative similarity along the diagonal $\delta=\epsilon$ in all three cases,
which is relevant for our definition of quantum Ricci curvature. Fig.\ \ref{fig:diag} shows the behaviour along the diagonal
of the sphere distance and the average sphere distance for all three two-dimensional model spaces, while Fig.\ \ref{fig:diagnorm} compares
the corresponding normalized distances, where we have divided by $\delta$. Again, the graphs illustrate the similarities
in behaviour of the sphere and the average sphere distances. One feature of the latter is that the three curves are genuinely
disjoint for $\delta >0$.
For the homogeneous and isotropic spaces we are considering presently, we can simplify the scaling relation (\ref{qric}) to
\begin{equation}
\frac{\bar{d}(S_p^{\delta},S_{p'}^{\delta})}{\delta}=c_q (1 - K_q(\delta)),
\label{qricsimp}
\end{equation}
We observe here that 
the constant $c_q$ can be determined uniquely and 
assumes the same value $c_q\approx 1.5746$ for all of the three smooth model spaces 
in two dimensions. 
Furthermore, we have established that the ``quantum Ricci curvature" $K_q(\delta)$ vanishes on flat space, and has a nontrivial
behaviour on the spaces of constant positive and negative curvature (Fig.\ \ref{fig:diagnorm}). It is negative and monotonically decreasing
on the negatively curved space, and is positive and monotonically increasing up to $\delta/\rho\approx 2.095$ on the positively curved
space. 

The two curves pertaining to the hyperbolic case in Fig.\ \ref{fig:diagnorm} both asymptote to 3, as can also be 
established straightforwardly by considering the limit $\delta\rightarrow \infty$ of the expressions for $d/\delta$ and $\bar{d}/\delta$ 
in eqs.\ (\ref{hyper2d}) and (\ref{hyp2dav}).
The plots on the right in both Figs.\ \ref{fig:diag} and \ref{fig:diagnorm} will serve as reference when we determine the curvature
properties of more general spaces.

\section{Curvature on regular lattices}
\label{regular:sec}

In this and the next section we will develop a better understanding of the behaviour of the quantum Ricci curvature on 
continuous but non-smooth metric
spaces. Since the eventual application we have in mind are Causal Dynamical Triangulations, we will focus on piecewise flat
spaces consisting of a single type of equilateral building block. In such a setting, the evaluation of the curvature assumes a combinatorial
character, because volume measurements reduce to a counting of building blocks (which all have equal size), and measuring
the geodesic (link) distance between two vertices $v_1$ and $v_2$ by definition amounts to a counting of one-dimensional edges in  
the shortest path linking $v_1$ and $v_2$. We treat these spaces as approximations to smooth spaces and therefore will be particularly
interested in the behaviour of curvature on scales that are large compared to the length $a$ of an edge of a building block. 
An important part of our analysis will be to obtain an estimate of the scale $\delta$ above which short-scale, so-called ``lattice artefacts" become irrelevant,
by which one means a dependence of the results on the specifics of the shape of the individual building blocks and of the local ``gluing rules" by
which the metric spaces are assembled from them.

The spaces we investigate in this section are flat, regular lattices, and can be regarded as tilings or subdivisions
into equal building blocks of flat, Euclidean space. We will treat the square, hexagonal and honeycomb lattices in two
dimensions and the cubic and face-centred cubic lattices in three dimensions.
To determine their quantum Ricci curvature, we will use a straightforward implementation of the average sphere
distance (\ref{sdist}) on these spaces, which is given
by formula (\ref{simpdist}) for two overlapping spheres $S^\delta$ whose radii are equal to the distance between
their centres, $\epsilon=\delta$. Some of the calculations are sufficiently simple to be done analytically, as we will see below.

To illustrate what is involved computationally, we will first consider the two-dimensional square lattice. Its vertices are
all points with integer coordinates $(x,y)$, and the geodesic link distance between two such points is the number of
edges of the shortest path between them.  
\begin{figure}[t]
\centerline{\scalebox{0.4}{\rotatebox{0}{\includegraphics{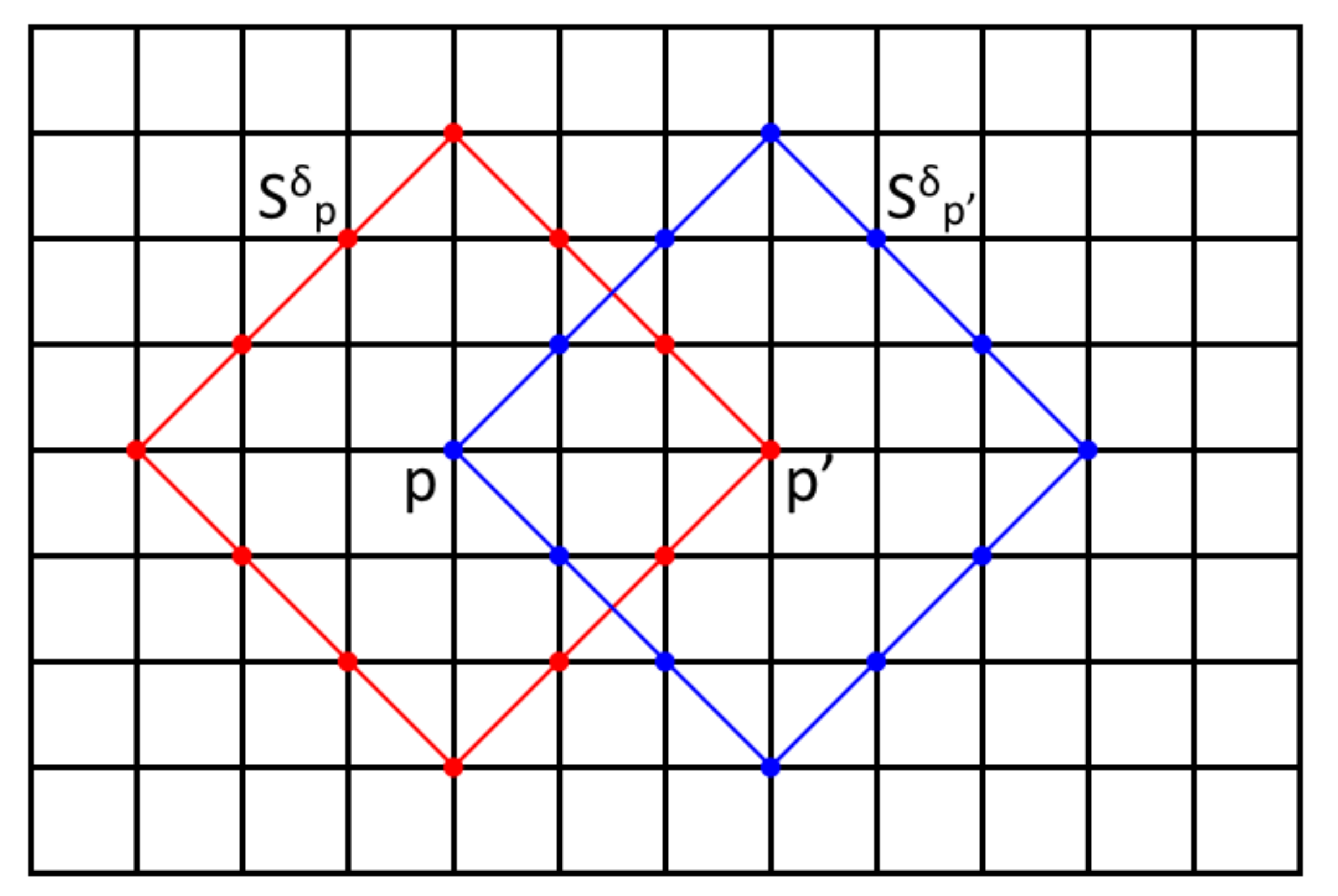}}}}
\caption{Two overlapping spheres $S^\delta_p$ and $S^\delta_{p'}$ of radius $\delta\! =\! 3$ on a square lattice,
whose centre vertices $p$ and $p'$ are a link distance $\delta\! =\! 3$ apart. Each sphere consists of 12 vertices. The diagonal
edges between the sphere vertices are drawn for ease of visualization only and are not part of the spheres or the underlying
lattice.}
\label{fig:square}
\end{figure}
Fig.\ \ref{fig:square} shows a pair of overlapping $\delta$-spheres, whose average distance one wants to compute as a function of
the scale $\delta$. Since the set-up is invariant under discrete lattice translation in both the $x$- and $y$-directions, one can
without loss of generality put the centre of the sphere $S^\delta_p$ at the origin, such that $p=(0,0)$. 
Note that the link distance between two points $p=(x,y)$ and $p'=(x',y')$ is given by
\begin{equation}
d(p,p')=|x-x'|+|y-y'|.
\label{latticedist}
\end{equation}
As an intermediate step to computing the sphere distance, one can work out the distance of an arbitrary point $(x,y)$ to 
the $\delta$-sphere $S^\delta_0$ centred at $(0,0)$, defined as $d(S^\delta_0,p):=\sum_{q\in S^\delta_0}d(q,p)$.  
Because of the lattice symmetries, it is sufficient to compute this distance for a point $p=(x,y)$ lying in the positive quadrant,
where $x\geq 0$ and $y\geq 0$. Distinguishing between four different cases, depending on the location of $p$, one finds
\begin{equation}
d(S^\delta_0,p)=\left\{\begin{array}{cl} 
4\delta (x+y),\;\;\; & x\geq\delta,\, y\geq\delta \\ 
4\delta x+2\delta^2+2 y^2,\;\;\; & x\geq\delta,\, 0\leq y<\delta\\
4\delta y+2\delta^2+2 x^2,\;\;\; & 0\leq x<\delta,\, y\geq\delta \\
4\delta^2+2 x^2+2 y^2,\;\;\; & 0\leq x<\delta,\, 0\leq y <\delta.
\end{array}\right.
\label{casesdist}
\end{equation}
If the centres of the two spheres share the same $x$- or the same $y$-coordinate, the shortest path between their
centres is a straight line segment, as in the example shown in Fig.\ \ref{fig:square}. Taking into account that the
volume of a $\delta$-sphere is given by $4\delta$ (the number of vertices contained in the sphere of radius $\delta$),
one obtains for the average sphere distance in this particular case
\begin{equation}
\bar{d}(S^\delta_p,S^\delta_{p'})=\frac{7}{4}\,\delta+\frac{1}{8\, \delta}=1.75\, \delta +0.125\, \frac{1}{\delta}.
\label{distsquaresphere}
\end{equation}
Recall that in the continuum case of the previous section, the corresponding expression for the flat case had a 
term linear in $\delta$, and no higher-order terms. Eq. (\ref{distsquaresphere}) for the square lattice 
has the same features, but with an additional term proportional to $1/\delta$. 
It will be suppressed with increasing $\delta$ and therefore can be interpreted as a short-scale lattice discretization effect. 
On the hexagonal lattice, which consists of equilateral triangles, the analogous scaling relation turns out to be
\begin{equation}
\bar{d}(S^\delta_p,S^\delta_{p'})=\frac{44}{27}\,\delta+\frac{1}{27\, \delta}\approx 1.6296\, \delta+0.0370\, \frac{1}{\delta},
\label{disthexa}
\end{equation}
where again we have considered overlapping spheres whose centres are connected by a straight sequence of edges.
The fact that the coefficients of the linear terms in (\ref{distsquaresphere}) and (\ref{disthexa})
differ from the continuum value of 1.5746 is due to two effects. First, the
use of link distance instead of Euclidean distance leads to an overestimation of distances except those along 
straight sequences of links. Second, the shape of a ``sphere" on a square or hexagonal lattice differs significantly
from that of a round sphere, which affects results. The fact that a hexagon is closer in shape to a sphere may
explain that the coefficient is closer to the continuum value. 

\begin{figure}[t]
\centerline{\scalebox{0.7}{\rotatebox{0}{\includegraphics{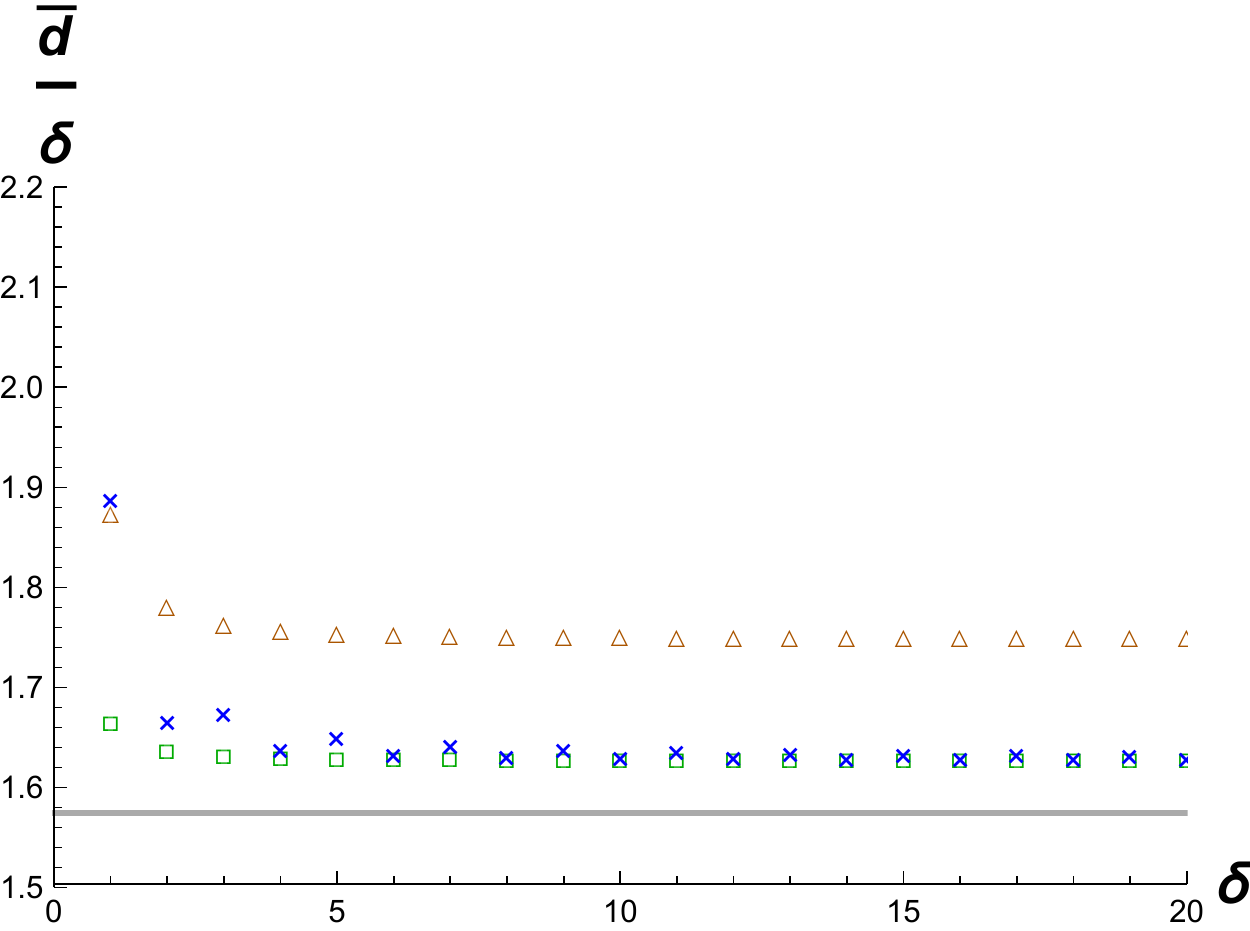}}}}
\caption{Normalized average sphere distance $\bar{d}(S^\delta_p,S^\delta_{p'})/\delta$ 
on the square, hexagonal and honeycomb lattices in two dimensions,
marked by triangles, squares and crosses respectively, as function of $\delta$.
The straight horizontal line is that of flat continuum space, and is included for comparison.}
\label{fig:2dregular}
\end{figure}
We have also investigated the honeycomb lattice. By eliminating every other vertex from it -- keeping only vertices
whose pairwise link distance is even -- one obtains a hexagonal lattice. This implies that the results for the 
average sphere distance on the honeycomb lattice for even $\delta$ will be the same as two times those for the
hexagonal lattice for $\delta/2$. The case of odd $\delta$ is slightly more involved and can be treated separately.
The complete result for the honeycomb lattice is given by
\begin{equation}
\bar{d}(S^\delta_p,S^\delta_{p'})=\Bigg\{\begin{array}{cl} 
\frac{44}{27}\,\delta+\frac{1}{18}+\frac{7}{27\,\delta}-\frac{1}{18\,\delta^2} &\;\; \delta\; {\rm odd},\\
\vspace{-0.3cm}&\\
\frac{44}{27}\,\delta+\frac{4}{27\,\delta} &\;\;\delta\; {\rm even} .
\end{array} 
\label{disthoney}
\end{equation}
Fig.\ \ref{fig:2dregular} shows the plots for the normalized average sphere distance $\bar{d}/\delta$ for
the three flat lattices. We observe that in all cases the curves start out at slightly elevated values for
small $\delta$ and then quickly settle down to a constant, as one would expect from a flat-space
behaviour, where the value of the constant depends on the lattice chosen. These differences are to be
expected, because on large scales the geodesic link distance scales with a different constant relative to
the ``true" geodesic distance in the continuum, depending on the type of lattice. If one wanted to take
the short-scale geometry of these lattices seriously, one would say that they exhibit negative quantum Ricci curvature 
for small $\delta$. Note that this phenomenon also occurs for the coarse Ollivier-Ricci curvature, which is
negative when evaluated at $\delta=1$ on a regular honeycomb lattice, say \cite{loiselromon}.

\begin{figure}[t]
\centerline{\scalebox{0.7}{\rotatebox{0}{\includegraphics{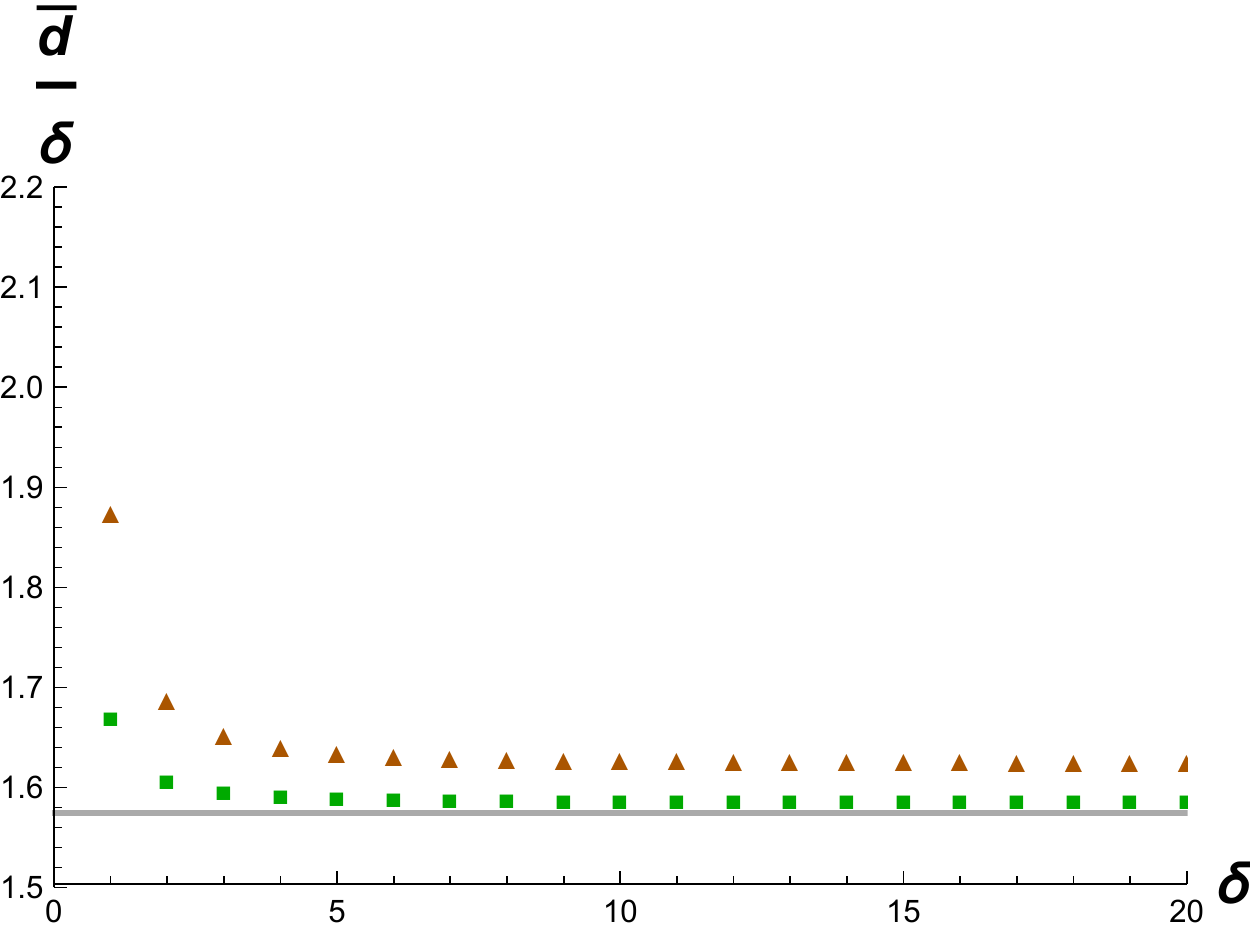}}}}
\caption{Normalized average sphere distance $\bar{d}(S^\delta_p,S^\delta_{p'})/\delta$
on two-dimensional flat lattices, 
averaged over lattice directions as described in the text.
Triangles and squares mark the data points for the square and hexagonal lattices
respectively.
The straight horizontal line is that of flat continuum space.}
\label{fig:2dregularav}
\end{figure}
As mentioned above, our computations for the average sphere distance did not use
the most general configuration of two overlapping spheres at distance $\delta$, but only 
pairs of spheres whose centres are aligned along a straight line. 
In the earlier example of the
square lattice depicted in Fig.\ \ref{fig:square}, these would be pairs of spheres whose centres
share the same $x$-coordinate, $p=(0,0)$ and $p'=(0,\delta)$, or the same $y$-coordinate, with
$p=(0,0)$ and $p'=(\delta,0)$. For the square and hexagonal lattices, we have repeated the calculation
of $\bar d$ for the most general case, where the shortest path connecting the two centres can
be a zigzag path.\footnote{On the honeycomb lattice, there are no straight paths in the sense of
Euclidean flat space, but there is an analogue of the preferred ``straight" directions of the
other lattice types, which was used to compute the formulas (\ref{disthoney}).} On the square lattice, 
this would be the case for centre coordinates
$p=(0,0)$ and $p'=(x,y)$ with $0<x<\delta$, $0<y<\delta$ and $x+y=\delta$, say. 
For a given distance $\delta$, there are now many more sphere configurations to consider and
an analytic evaluation is less straightforward. Instead, we have used the computer to calculate
$\bar{d}/\delta$ based on this more general set of configurations. One would a priori expect
that the underlying improved averaging over lattice directions leads to results closer to those of
the continuum. This is indeed the case, as illustrated by the data shown in Fig.\ \ref{fig:2dregularav}. 
The convergence behaviour is similar to that depicted in Fig.\ \ref{fig:2dregular}, but the
constant asymptotic values 1.625 for the square lattice and 1.583 for the hexagonal lattice are
closer to the value found in the continuum. 

Turning finally to three-dimensional lattices, 
a similar derivation for the flat cubic lattice leads to an average
sphere distance 
\begin{equation}
\bar{d}(S^\delta_p,S^\delta_{p'})=\frac{82\delta^5+90\delta^3+23\delta}{10(1+2\delta^2)^2}=
\frac{41}{20}\,\delta+\frac{1}{5\, \delta} +{\cal O}((\tfrac{1}{\delta})^3),
\label{distcubic}
\end{equation}
where again we have considered only those configurations where the centres of the two overlapping two-spheres 
are separated by a straight sequence of $\delta$ lattice edges. To determine the distance between
the spheres, we averaged over the distances between all pairs of vertices contained in the two spheres. 
The result (\ref{distcubic}) strongly resembles the behaviour in two
dimensions, with an asymptotically linear behaviour in $\delta$ and positive ``correction terms" for small $\delta$.
Again the coefficient of the linear term, $41/20=2.05$ is larger than the corresponding continuum value 1.6250.

We have also investigated the face-centred cubic lattice, which is associated with a closest packing of
spheres in three dimensions. To construct it, one starts with a single layer of spheres, arranged in a closest packing with
respect to two dimensions, the $x$-$y$-plane, say.
The centres of the spheres can be thought of as the vertices of a two-dimensional
lattice, whose edges correspond to pairs of neighbouring spheres. Since each sphere has six neighbours,
this results in the two-dimensional regular hexagonal lattice we already discussed above. On top of the
lowest layer, we stack another, identical layer of spheres in the $z$-direction. Since there are twice as many gaps
in the lower layer as there are spheres in the second layer, there are two possibilities of placing 
the second layer, corresponding to two different displacements of the spheres relative to
those of the first layer. Each sphere in the lower layer has three neighbouring
spheres in the second layer, and vice versa. There are two different choices for how to add
a third layer of spheres. The first possibility is to align the centres of the spheres in the $x$-$y$-directions
with those of the first layer, and the second possibility -- the one chosen by us -- is to displace the centres in
the same direction and by the same amount in the $x$-$y$-plane as in the step from the first to the second layer.   
Repeating the same step for subsequent layers, one obtains a regular three-dimensional lattice with discrete period
3 in the $z$-direction, the so-called face-centred cubic lattice, all of whose vertices have order 12. 

We were able to derive an explicit formula for the average
sphere distance on this lattice, for the case that the centres of the spheres lie in the same hexagonal layer  
and are connected by a straight sequence of lattice edges. The result is given by
\begin{equation}
\bar{d}(S^\delta_p,S^\delta_{p'})=\frac{3547\delta^5+1705\delta^3+148\delta}{80(1+5\delta^2)^2}=
\frac{3547}{2000}\,\delta+\frac{1431}{10000\, \delta} +{\cal O}((\tfrac{1}{\delta})^3),
\label{distfcc}
\end{equation}
and therefore structurally similar to the result for the cubic lattice, eq.\ (\ref{distcubic}). 
The coefficient of the linear term is $3547/2000\approx 1.77$, which is closer to the continuum value than
that of the cubic lattice. This resembles the situation we encountered in two dimensions, namely,
that the lattice with the higher coordination number (in this case 12 instead of 8) 
appears to give a better approximation to the continuum.

To summarize, evaluating the average sphere distance on several flat regular lattices, viewed as discrete approximations
to continuum flat space, leads to consistent results: up to short-distance lattice artefacts, which are confined to
a region $\delta \lesssim 5$, the behaviour of $\bar d$ is essentially linear in $\delta$, compatible 
with a vanishing quantum Ricci curvature $K_q(\delta)$ in eq.\ (\ref{qricsimp}). In all the cases we have investigated,
the constant $c_q$ in the same scaling law is in the vicinity of and larger than the corresponding continuum value.

\section{Curvature on random triangulations}
\label{random:sec}

As a next step we consider classes of random triangulations that in general carry nonvanishing quantum Ricci curvature. 
They are still well-behaved in the sense of not deviating too much from smooth spaces. The triangulations
are two-dimensional, made of equilateral Euclidean triangles and are obtained from Delaunay triangulations of flat and
constantly curved spaces of either signature. Their small-scale behaviour
depends on the local random structure, but their properties on large scales reflect the geometry of the smooth spaces
they are approximating. 

\begin{figure}[t]
\centerline{\scalebox{0.45}{\rotatebox{0}{\includegraphics{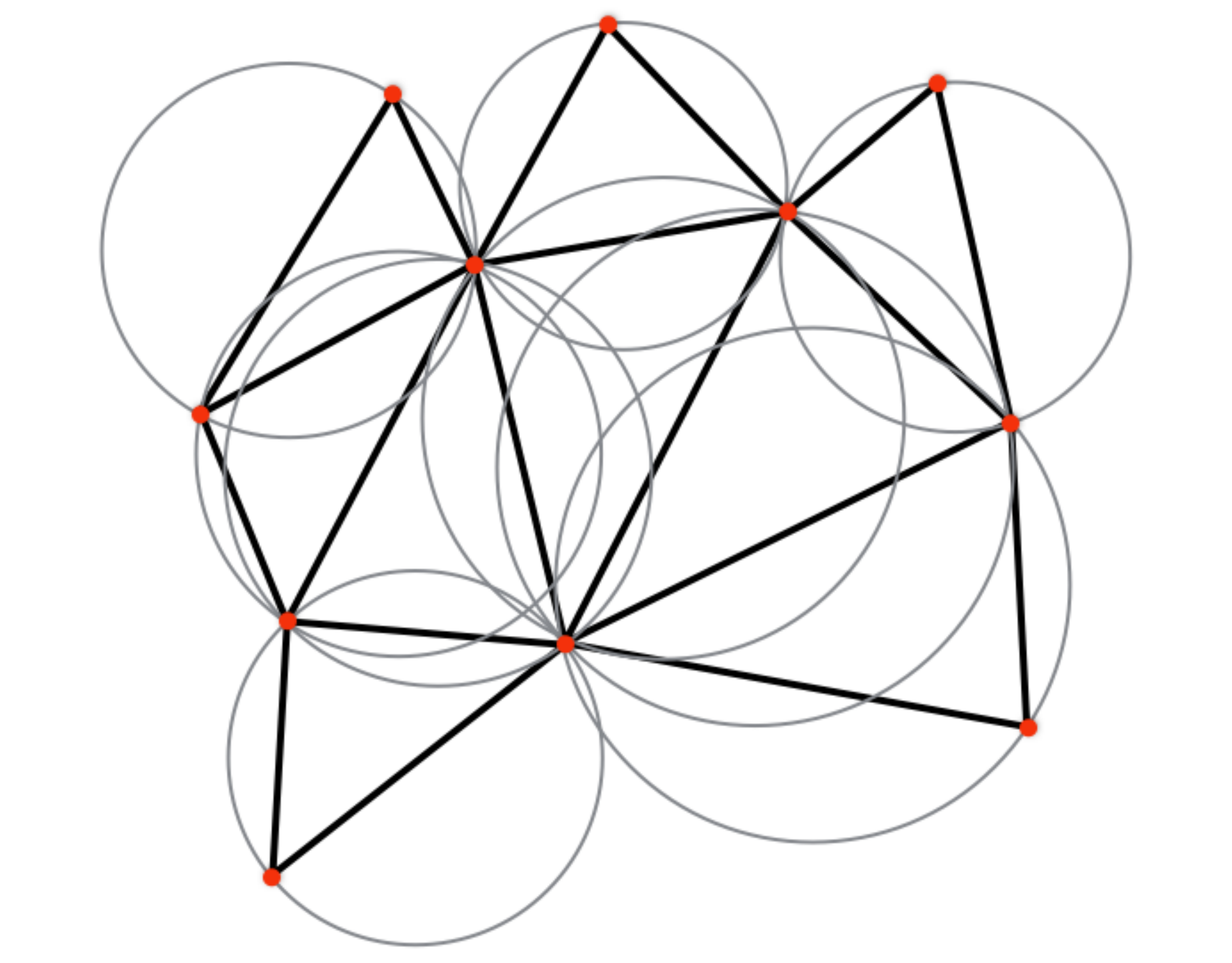}}}}
\caption{A Delaunay triangulation in the plane, together with the circumcircles of its constituting triangles.
By definition, no circumcircle contains any vertices of the triangulation in its interior.}
\label{fig:delaunay}
\end{figure}

Recall that a Delaunay triangulation in the plane is a
triangulation of a finite point set $P\subset \R^2$ (constituting the vertices of the triangulation) 
if the circumcircle of every triangle contains no points of $P$ in its interior. The circumcircle of a triangle
is defined as the unique circle containing the three vertices of the triangle (see Fig.\ \ref{fig:delaunay}). 
Because of their
nice geometric properties Delaunay triangulations appear in numerous applications. 
Compared to other triangulations of the same point set $P\subset\R^2$, the 
(essentially unique\footnote{The uniqueness is up to sets of more than $D+1$ vertices (in $D$ dimensions) that fall on the same
circle, without other vertices inside the circle. For a local configuration of this type, any valid internal substructure will lead
to a Delaunay triangulation. In our construction, this degeneracy cannot occur.}) Delaunay triangulation of 
$P$ maximizes the minimum angle, which means that 
thin, elongated triangles tend to be avoided. Note that analogous constructions of Delaunay triangulations exist
in higher dimensions too.

In all cases, we will proceed in three steps, first generating a point set $P$ with the help of a Poisson disc sampling. 
Poisson disc sampling generates a
tightly packed point collection with a specified minimal distance $d_{\rm min}$ between any two of its points. 
Second, we construct a Delaunay triangulation that has these points as vertices. 
Because of the nature of the Poisson disc sampling, 
the edge lengths of this Delaunay triangulation are clustered relatively compactly around some average
edge length. 
The third step consists in setting all edge lengths to 1 and thereby making the triangulations equilateral before
starting to measure average sphere distances on them. This is motivated by the fact that we are interested in 
exploiting the simplicity of the combinatorial aspects of the prescription (\ref{simpdist}), which also holds in 
CDT quantum gravity, the physical application we are primarily interested in. Adjusting the edge lengths in this way
will alter the local metric properties of the triangulations. However, this appears to have only a mild effect,
which is confined to smaller
scales, as we will see when examining the results of the quantum Ricci curvature measurements. 

\subsection{Flat case}

We begin by sketching the procedure for the case of random triangulations approximating
flat space, where we will use an auxiliary regular grid to speed up the Poisson disc sampling, following \cite{poisson}. 
We refer the interested reader to reference \cite{poisson} for further details on the construction. 
For the process to be meaningful, we must confine ourselves to a finite region of flat space, which we choose to be
a square of approximate side length 100 $d_{\rm min}$. 
All subsequent operations will take place inside this square.\footnote{Since we have only treated 
the computation of sphere distances for interior points, we make sure that 
during measurements we stay well away from any boundaries. It would take us too far to give a detailed description
of the boundary construction for our triangulations. Suffice it to say that it involved a one-dimensional Poisson process 
with minimal distance $d_{\rm min}$, and that we performed detailed numerical tests to make sure unwanted boundary effects 
are negligible.} Furthermore, the square is overlaid by 
a regular square grid whose cells have side length $d_{\rm min}/\sqrt{2}$. This ensures that
each cell will contain at most one point of the point set $P$ to be constructed. The grid forms
an auxiliary structure in the Poisson disc sampling and the subsequent triangulation. 

\begin{figure}[t]
\centerline{\scalebox{0.4}{\rotatebox{0}{\includegraphics{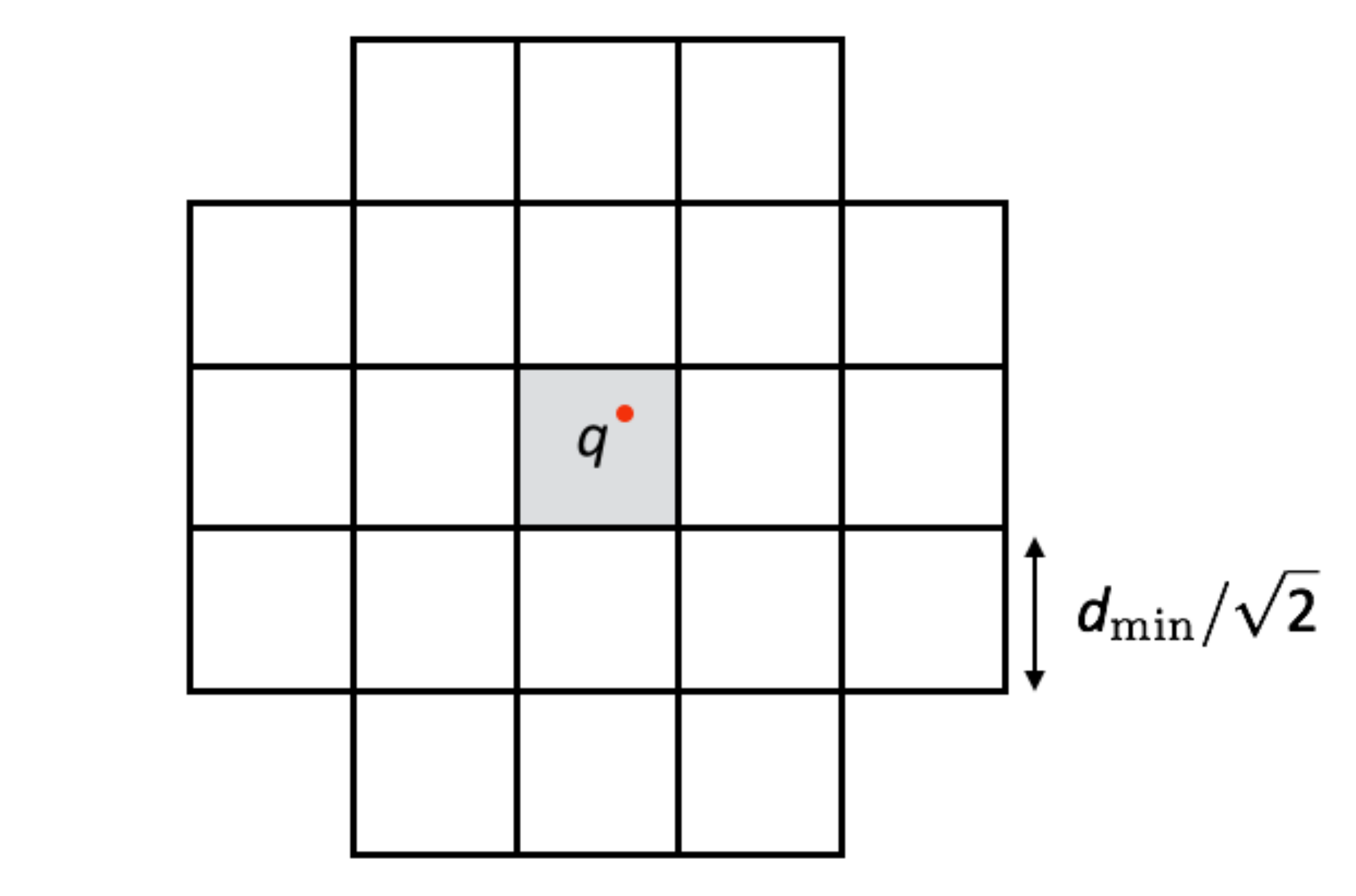}}}}
\caption{Using the auxiliary square lattice: during the construction of the initial point set $P$, only the 20 cells shown
surrounding the cell of a new candidate point $q$ need to be checked for the presence of other points within radius $d_{\rm min}$.}
\label{fig:pointgrid}
\end{figure}

Starting from an initial point $p_0$ at the centre of the square, say, we systematically build up a point set $P$.
The process is characterized by the minimal distance $d_{\rm min}$ and by another integer $k$, which is chosen
a priori and will determine the density of $P$. A step in the algorithm consists in picking a point $p$ 
from the set of points already selected to lie in $P$. Given $p$, we randomly pick a new point $q$ in the annulus between radii
$d_{\rm min}$ and $2\;\! d_{\rm min}$ around $p$. If the Euclidean distance of $q$ 
to any other already selected point is smaller than
or equal to $d_{\rm min}$, the point is discarded, otherwise it is added to the set of points selected to lie in $P$. For fixed
$p$, we generate $k$ new random points in this way, which we either keep or discard. 
A larger $k$ will lead to a denser and more uniform set $P$ at the end of the algorithm, but also to an increase in the overall 
time needed to generate the points. We used $k=30$. Next, another point $p'$ is taken from the already
selected point set and the procedure is repeated by choosing $k$ times a random point in the annulus around $p'$. 
Note that a point $p$ can only serve once as the base point for such a search, lying at the centre of an annulus. 
The process ends when all points in the selected set have acted as a base point. The final point set is the
searched-for $P$. Note that the presence of the grid structure simplifies the
test of whether a point $q$ should be discarded, because only a finite number of cells (20 cells excluding the cell where
$q$ itself is located) around $q$ need to be
checked for points that are potentially too close to $q$, see Fig.\ \ref{fig:pointgrid} for illustration. 

The same grid structure is also used during the construction of the Delaunay triangulation of a given set $P$. 
Following \cite{poisson}, we first generate a discrete clustering of all cells, where each cell $C$ is associated with the vertex in $P$
that is closest to the centre of $C$ in terms of Euclidean distance. This results in a clustering of the cells of the square grid,
with as many clusters as there are vertices in $P$. The algorithm proceeds by examining each vertex {\it of the square grid} in turn,
by picking for each cell its lower left-hand corner point, say. Each corner point $x$ is then assigned an index between 1 and 4, 
counting the number of distinct clusters meeting at $x$. The index is 1 if all four cells meeting at $x$ belong to the same cluster,
it is 2 if the four cells belong to two different clusters, and analogous for index 3 and 4, see \cite{poisson} for 
further explanations and illustrations. 

The point of this clustering is that it allows for the straightforward construction of a
triangulation that is ``almost Delaunay". 
To obtain it, we draw for each corner point with index 3 a triangle connecting the corresponding three vertices of $P$.
Next, we draw for each corner point with index 4 the quadrilateral spanned by the corresponding four vertices. There are then two
ways to add an interior link to obtain a pair of adjacent triangles. Of those, we choose the interior link for which the angle sum
of the quadrilateral at the corners met by the link is larger than $\pi$. This is a necessary condition for a Delaunay triangulation,
and equivalent to the circumcircle condition mentioned earlier.
After dealing with all corner points of index 3 and 4 in this manner, one obtains a triangulation which in general is not quite
a Delaunay triangulation, but can be transformed into one by systematically checking the local Delaunay property for every link,
and performing link flips wherever necessary, as shown in reference \cite{poisson}.

\subsection{Non-flat case}

The procedure outlined in the previous subsection must be adapted for random triangulations approximating non-flat
spaces. The first step will again be to construct a point set $P$ by Poisson disc sampling, this time on a constantly curved,
smooth model space, the two-dimensional sphere or (a subset of) two-dimensional hyperbolic space. In both cases
we have found it convenient to parametrize points in these spaces by the Cartesian coordinates $(x,y,z)$ of their
embeddings into $\R^3$, as described in Sec.\ \ref{smooth:sec} above. 
Introducing the notation 
\begin{equation}
q_1\cdot q_2\equiv (x_1,y_1,z_1)\cdot (x_2,y_2,z_2)=x_1x_2+y_1y_2+z_1z_2
\label{scalareu}
\end{equation}
for the scalar product (the flat Euclidean metric) of elements $q_i\in\R^3$, recall that we
defined the two-sphere as all points $q$ with $q\cdot q=\rho^2$. On this two-sphere,
the flat metric (\ref{scalareu}) induces a constantly curved metric with curvature
$+1/\rho^2$. All distance measurements, including those occurring during the Poisson disc sampling on
the sphere, have to be done with respect to this nontrivial metric.  
Similarly, introducing the notation 
\begin{equation}
q_1\ast q_2\equiv (x_1,y_1,z_1)\ast (x_2,y_2,z_2)=x_1x_2+y_1y_2-z_1z_2
\label{scalarmin}
\end{equation}
for the indefinite scalar product (the flat three-dimensional Minkowski metric) for elements $q_i\in\R^3$,
we define hyperbolic space as all points $q$ for which $q\ast q=-\rho^2$ and $z>0$.
On this upper sheet of the two-dimensional hyperboloid, the metric (\ref{scalarmin}) induces a constantly curved, positive definite metric with 
curvature $-1/\rho^2$. Again, this nontrivial metric must be used when measuring geodesic distances on the two-dimensional
hyperbolic space. 

Note that for a pair of points $(p,q)$ on the two-sphere, given in terms of their Cartesian coordinates, their geodesic distance
on the sphere can be expressed with the help of the scalar product (\ref{scalareu}) as
\begin{equation}
d(p,q)=\rho \arccos\left(\frac{p\cdot q}{\rho^2}\right).
\label{disteu}
\end{equation}
In a similar fashion, the geodesic distance of two points $(p,q)$ on hyperbolic space, given in terms of their Cartesian coordinates is
\begin{equation}
d(p,q)=\rho\, {\rm arccosh}\left(-\frac{p\ast q}{\rho^2}\right),
\label{distmin}
\end{equation}
using the inner product (\ref{scalarmin}). The minus sign in the argument of the inverse hyperbolic cosine comes from our choice 
of overall sign in the Minkowskian scalar product (\ref{scalarmin}). Relevant for the construction of an annulus around
some point $p$ on the two-sphere -- needed in the Poisson disc sampling -- is the fact that the set of all points at a constant distance
$d$ from $p$ {\it on the sphere} also forms a planar circle in the embedding space $\R^3$. This is made explicit by expressing the scalar product
$p\cdot q$ in eq.\ (\ref{disteu}) in terms of the {\it three}-dimensional Euclidean distance $d_{\rm eu}(p,q)$ of the two points, leading to
\begin{equation}
d_{\rm eu}(p,q)=2\rho\sin\left( \frac{d(p,q)}{2\rho}\right).
\label{screlate}
\end{equation} 
This is an injective relation between $d$ and $d_{\rm eu}$ as long as $d<\pi\rho$, a condition that 
in our applications was always satisfied. 

When implementing the
Poisson disc sampling on the sphere, after picking a point $p$ to serve as the centre of an annulus, we 
parametrize the neighbourhood of $p$ in terms of a two-dimensional system of radial coordinates $(r,\varphi)$ centred
at $p$, such that the inner and outer boundary of the annulus (at geodesic distances $d_{\rm min}$ and $2\;\! d_{\rm min}$)
are circles of constant radius $r$. Like in the flat case, points in the annulus are then created randomly and uniformly, this time with
respect to the appropriate measure on the two-sphere, expressed in terms of the variables $r$ and $\varphi$.    
For each newly created point, we perform a test to make sure that its distance to all other points already included in the set $P$ 
is larger than $d_{\rm min}$. If this is the case, the point is added to the set, otherwise it is discarded. 
We did not attempt to set up suitable analogues of the square grid on the sphere or the hyperboloid to speed up this part of the
algorithm, and instead simply computed the distance of a given candidate point to all other points. 
Since we considered only relatively small configurations with
up to 20.000 points, the resulting increase in computational complexity to $O(n^2)$ could be handled without problems. 

The Poisson disc sampling in the hyperbolic case proceeds along similar lines, the only minor difference being that
the set of all points equidistant to a given point $p$ on the hyperboloid do in general not lie on a circle with respect
to the {\it Euclidean} metric of the embedding $\R^3$. To nevertheless be able to use a straightforward generalization of the
procedure on flat space and the sphere, we boost the centre $p$ of an annulus to the lowest point $(0,0,\rho)$ of
the hyperboloid, because in this case the set of all points equidistant to $p$ does lie on a planar circle in the embedding space.
We can again introduce a spherical coordinate system on a local, two-dimensional neighbourhood of $p$ and implement the disc sampling 
as before, with respect to the induced, non-trivial measure on the hyperboloid. Once a candidate point has been chosen randomly from the annulus, it
is boosted back, after which the usual distance check to all other points is performed with the help of eq.\ (\ref{distmin}).

The next step consists in generating Delaunay triangulations from the point sets $P$ we have constructed on the curved spaces using
Poisson disc sampling, as described above. 
In the curved context, we again define a Delaunay triangulation through the property that any (geodesic) circumcircle of
the triangulation built from $P$ does not have any elements of $P$ inside. This construction remains meaningful -- in the sense of
resembling the procedure in flat space -- as long as the size of the triangles is small compared to the curvature radius of the 
constantly curved spaces we are considering, which was always the case. 

The code we used to generate the triangulations
is based on reference \cite{dirichlet}, which makes use of Voronoi diagrams (also called Voronoi or Dirichlet tessellations).
Recall that the Voronoi diagram associated with a finite point set $P$, for simplicity taken to lie in the Euclidean plane, 
partitions the plane into
cells. Each cell is associated with a point $p\in P$ and consists of all points in $\R^2$ that are closer to $p$ than to any other 
point of $P$, 
so that each cell has the shape of a convex polygon. The set of all line segments forming the borders between adjacent cells
forms a graph whose vertices are tri- or higher-valent. A generic point set, like the random sets $P$ 
we construct with the help of the Poisson disc sampling, has a unique, trivalent ``Voronoi graph" 
associated with it, which in turn is dual to the unique
Delaunay triangulation constructed from the same point set. An analogous construction also goes through for the ``mildly curved" spaces
we are considering, with the Euclidean distance substituted by the appropriate geodesic distance on these spaces. Note that 
the vertices of the Voronoi diagram coincide with the centres of the circumcircles of the dual Delaunay triangulation. 

The algorithm in \cite{dirichlet} proceeds iteratively, adding in each step a vertex to an already existing Delaunay triangulation.
Data are stored and manipulated referring to the vertices of the triangulation as well as to the (dual) vertices of the Voronoi
diagram, which also means that the new elements of the latter have to be computed and updated in each step. The beauty of
the set-up lies in the fact that these updates only affect small local neighbourhoods of the triangulation. We will not describe
details of the algorithm here, which can be found in \cite{dirichlet} for Euclidean spaces, but only describe the modifications 
that are necessary in the curved case.  
 
Firstly, we need to choose an initial Delaunay triangulation. For the case of positive curvature, we pick four vertices on the sphere which
span an equilateral tetrahedron in the embedding $\R^3$, and connect them by geodesic arcs on the sphere. Obviously, the length
of these initial edges is much larger than $d_{\rm min}$, but they quickly disappear as the algorithm progresses, since it includes
the creation and removal of links in each step. By contrast, for the case of negative curvature,
since the upper sheet of the hyperboloid has infinite
volume, we must impose a cutoff to make the construction well defined. Our prescription was to consider only points with
embedding space coordinate $z\leq z_{\rm max}= 3\rho$. Just like in flat space, we therefore are dealing with a spatial region
with a boundary. Vertices on the boundary $z=3$ were again generated with a one-dimensional Poisson sampling of geodesic
distances in the interval $[d_{\rm min},2\;\! d_{\rm min}]$. The initial triangulation of this hyperbolic disc is obtained by connecting
each boundary vertex to the apex $(0,0,\rho)$ of the hyperboloid by a geodesic line segment. The length of these segments
exceeds $2\;\! d_{\rm min}$, but again this does not seem to leave any imprint on the final triangulations.

Secondly, we need an effective method to compute the locations of the vertices of the Voronoi diagram dual to a given Delaunay
triangulation. More specifically, we must determine the centre of a circumcircle spanned by a triple of vertices of
the triangulation, which requires a simple application of linear algebra. Put briefly, for both the sphere 
and the hyperboloid one first
identifies the plane in $\R^3$ spanned by the difference vectors of the three vertices, using the cross product of vectors or
a Gram-Schmidt procedure respectively. One then looks for the axis through the origin 
in $\R^3$ which is perpendicular to that plane, using the inner products (\ref{scalareu}) or (\ref{scalarmin}) as appropriate,  
and finally determines the point in which the axis meets the sphere or the hyperboloid.

\subsection{Measurement method}
\label{measure:sec}

\begin{figure}[t]
\centerline{\scalebox{0.45}{\rotatebox{0}{\includegraphics{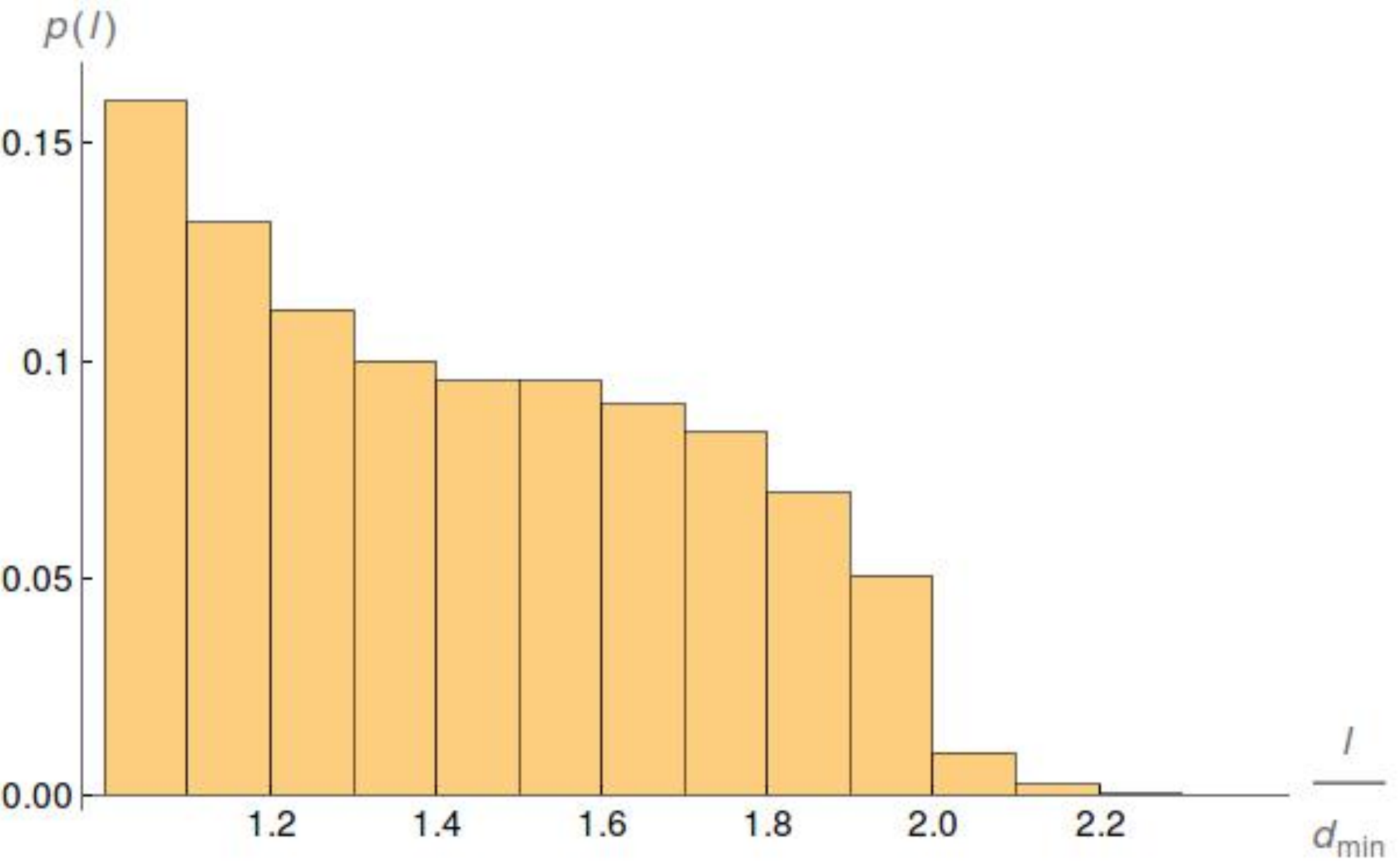}}}}
\caption{Probability distribution $p(\ell)$ of edge lengths $\ell$ in a Delaunay triangulation of flat space with 6.200 vertices, 
binned in intervals of 0.1 $d_{\rm min}$.
}
\label{fig:LDF}
\end{figure}

For all three types of geometry, the final step in constructing the triangulations that we will use for exploring our curvature
prescription is to set all edge lengths of the Delaunay triangulations to unity. To give a quantitative impression of 
the distribution of edge lengths $\ell$ before making the triangulation equilateral, Fig.\ \ref{fig:LDF} shows a sample from a 
Delaunay triangulation of flat space. The edge lengths are distributed rather evenly across the interval $[d_{\rm min},2\;\! d_{\rm min}]$, 
increasing somewhat in the vicinity of $d_{\rm min}$, which by construction constitutes a kinematical lower bound, and
decreasing towards longer lengths. 
The fact that very few edge lengths exceed $2\;\! d_{\rm min}$ reflects the well-behaved geometry of the
triangulation.

Before embarking on the curvature measurements, we measured some other geometric properties of the triangulations to
check whether they are roughly in line with those of the corresponding continuum geometries. 
For all three types of geometry, we measured the distribution of the vertex order\footnote{In two dimensions, the vertex order (the 
number of links meeting at a vertex) is
a direct measure of the deficit angle and therefore of the local Gaussian curvature at a vertex.},
and also determined the scaling of the size 
of geodesic spheres (circles in our case) as function of their geodesic radius, and compared it to the
corresponding continuum behaviour. For the spherical case, we also measured the distribution of diameters, 
where the diameter at a vertex is defined as the distance to the furthest vertex in the triangulation. 
By and large, these quantities behave as expected from a comparison with their continuum counterparts, 
as will be discussed below. This indicates that the configurations continue to be ``nice"  and compatible with 
an overall spatial uniformity also after removing 
the differences between length assignments from the Delaunay triangulations.

We then collected data on the average sphere distances $\bar{d}(S^\delta_p,S^\delta_{p'})$ as a
function of the geodesic (integer) link distance $\delta$ in the range $\delta\in [1,15]$ for a given type
of geometry (flat, spherical or hyperbolic),
by averaging in each case over a set of ten
independent triangulations, and over the location and relative
orientation of pairs of spheres $S^\delta_p$ and $S^\delta_{p'}$. 

For a given triangulated configuration, the latter averages were implemented as follows.
After picking a vertex $p$ in the triangulation, we constructed its $\delta$-sphere $S^\delta_{p}$, 
consisting of all vertices at link distance $\delta$ from $p$, and determined the total number of vertices 
in the sphere. For each of the vertices $p'\in S^\delta_{p}$ in turn, we then constructed a new $\delta$-sphere
$S^\delta_{p'}$ centred at $p'$ and measured the average sphere distance $\bar{d}(S^\delta_p,S^\delta_{p'})$.
Averaging the resulting data over $p'$ for given $p$ implies an averaging over directions around $p$ on
the underlying space, thus removing directional information and producing an effective Ricci scalar curvature. 
Since we modelled our triangulations on isotropic continuum spaces, we expect them to be (approximately) isotropic
too. Averaging over directions in this case is trivial and will just contribute to reducing numerical errors.

The way we picked a sequence of initial points $p$ for a given configuration, for which the set of measurements just
described was performed for all $\delta\leq 15$, was by simply using the first 20 points that were created during the
Poisson disc sampling for this geometry. Recall that for the flat and hyperbolic spaces, which both have a boundary, 
we chose the initial point for the disc sampling to coincide with the centre of the space. Since points generated subsequently
always lie within an annulus of a previously generated point, this implies that the first 20 points from
such a sequence will be clustered not too far away from the centre. This was done mainly to avoid that the
measurements run into the boundary of the triangulation.\footnote{We always made sure by additional checks that
the minimal distance to the boundary of any point $p$ was larger than $2\delta$.} It also means that our measurements will
inevitably have some spatial overlap, and therefore not all data will be independent. However, since we also
averaged over different configurations, we do not think that this procedure leads to any systematic errors.  

\begin{figure}[t]
\centerline{\scalebox{0.55}{\rotatebox{0}{\includegraphics{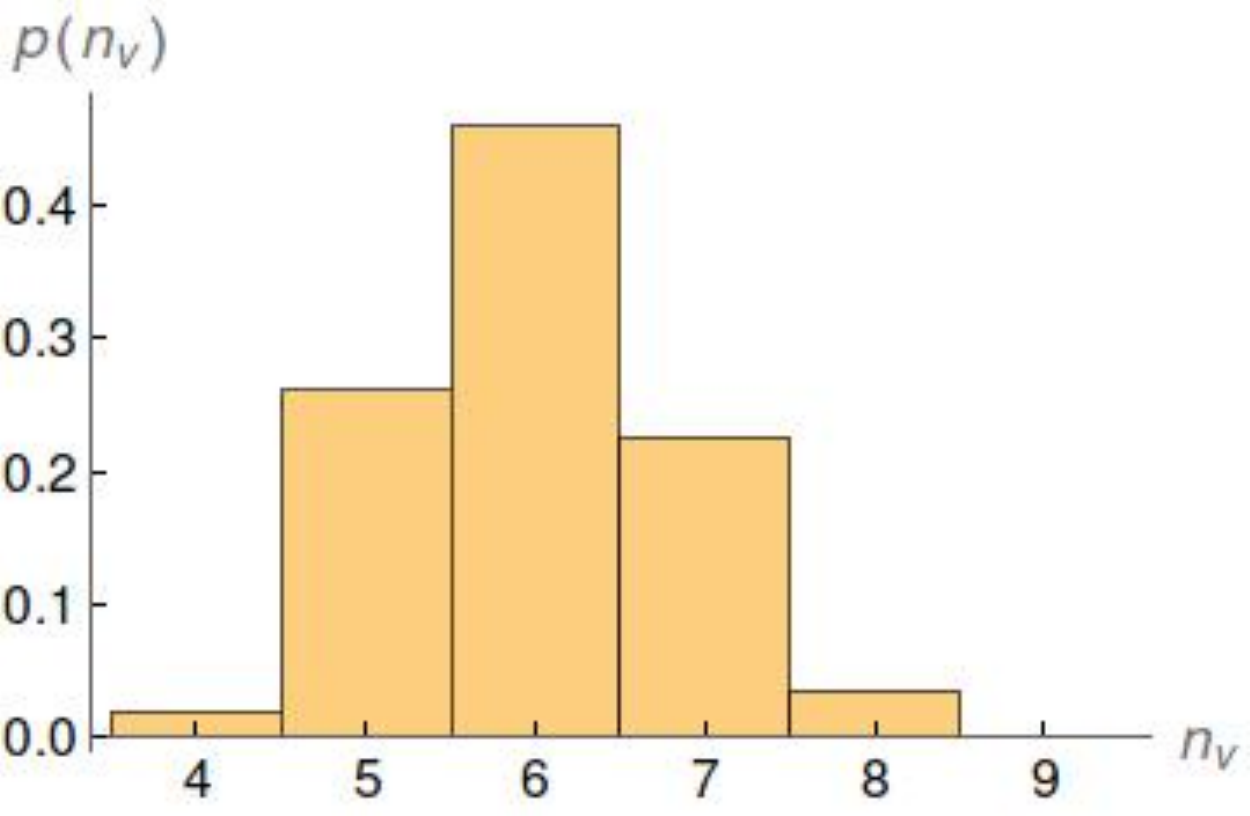}}}}
\caption{Distribution $p (n_v)$ of the vertex order $n_v$ of interior vertices of a Delaunay triangulation 
of a piece of flat space with 6.200 vertices.
}
\label{fig:vertexdist}
\end{figure}

\subsection{Measurement results}
\label{measureresults:sec}

Starting with the flat case, we investigated ten independent configurations, each with approximately 6.200 vertices. 
Fig.\ \ref{fig:vertexdist} shows
the distribution of the order $n_v$ of interior vertices of a sample triangulation. It is
centred around 6, with more than 90\% of vertices having coordination number 5, 6 or 7. The construction makes it
impossible to have internal vertices of order smaller than 4, which explains why 4 is the lowest order observed. In
the measurements considered we did not encounter vertices whose order was above 10. This is different from what 
happens in quantum configurations, like those appearing in dynamical triangulations, where the order distribution typically
has a long tail at high vertex orders. The absence of this feature for the Delaunay triangulations is another indicator of
their well-behaved nature.

\begin{figure}[t]
\centerline{\scalebox{0.7}{\rotatebox{0}{\includegraphics{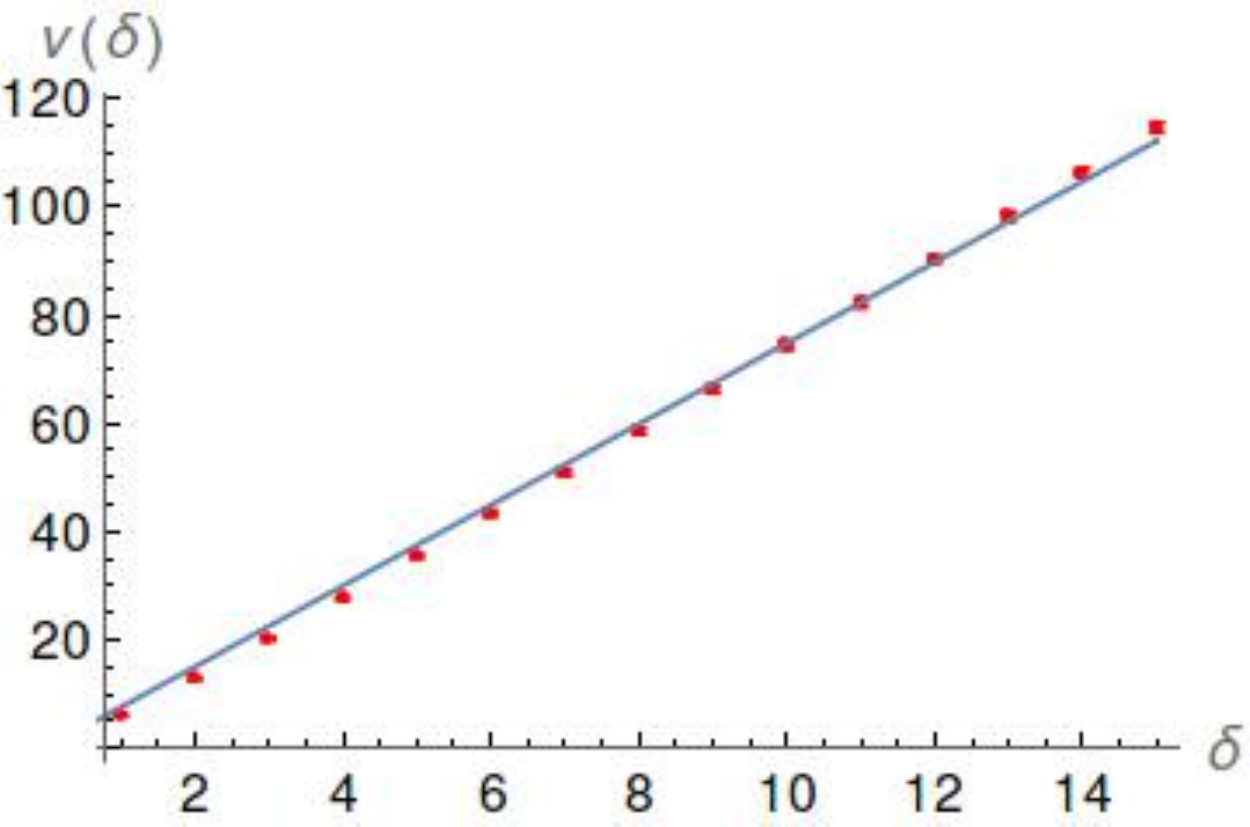}}}}
\caption{The (averaged) size $\nu(\delta)$ of circles as a function of their radius $\delta$, on a geometry obtained by 
setting the edge lengths of
a flat Delaunay triangulation to unity, including the best linear fit.
}
\label{fig:spherevolflat}
\end{figure}

A first check of the flat character of the triangulations is a measurement of the scaling behaviour of geodesic circles,
as explained in the previous subsection. We will denote the (discrete) volume of a circle of geodesic radius 
$\delta$ -- equal to the number of vertices contained in the circle -- by $\nu (\delta)$. A linear dependence on $\delta$
indicates flat-space behaviour, the corresponding relation in the continuum being $\nu (\delta)=2\pi\delta$. 
In a context where distances
are discretized because of the presence of building blocks of standard size, the proportionality constant on the
right-hand side of this equation will typically not be equal to $2\pi$, as is illustrated by the two-dimensional hexagonal
lattice, one of the regular lattices we explored in Sec.\ \ref{regular:sec}, for which we have $\nu(\delta)\! =\! 6\delta$. This
is a consequence of the lattice structure, where geodesic distances are not measured along straight lines in the conventional
continuum sense, and where geodesic spheres are not smooth objects either. The data 
for the circle volume $\nu (\delta)$ collected from the ten configurations are displayed in
Fig.\ \ref{fig:spherevolflat}, together with a best linear fit for the average circle volume, given by $\nu (\delta)= 7.48(5)\delta$. 
In the $\delta$-range considered, the quality of the fit is good, showing that the behaviour is compatible with that
of a flat space on scales sufficiently large relative to the lattice spacing.

\begin{figure}[t]
\centerline{\scalebox{0.6}{\rotatebox{0}{\includegraphics{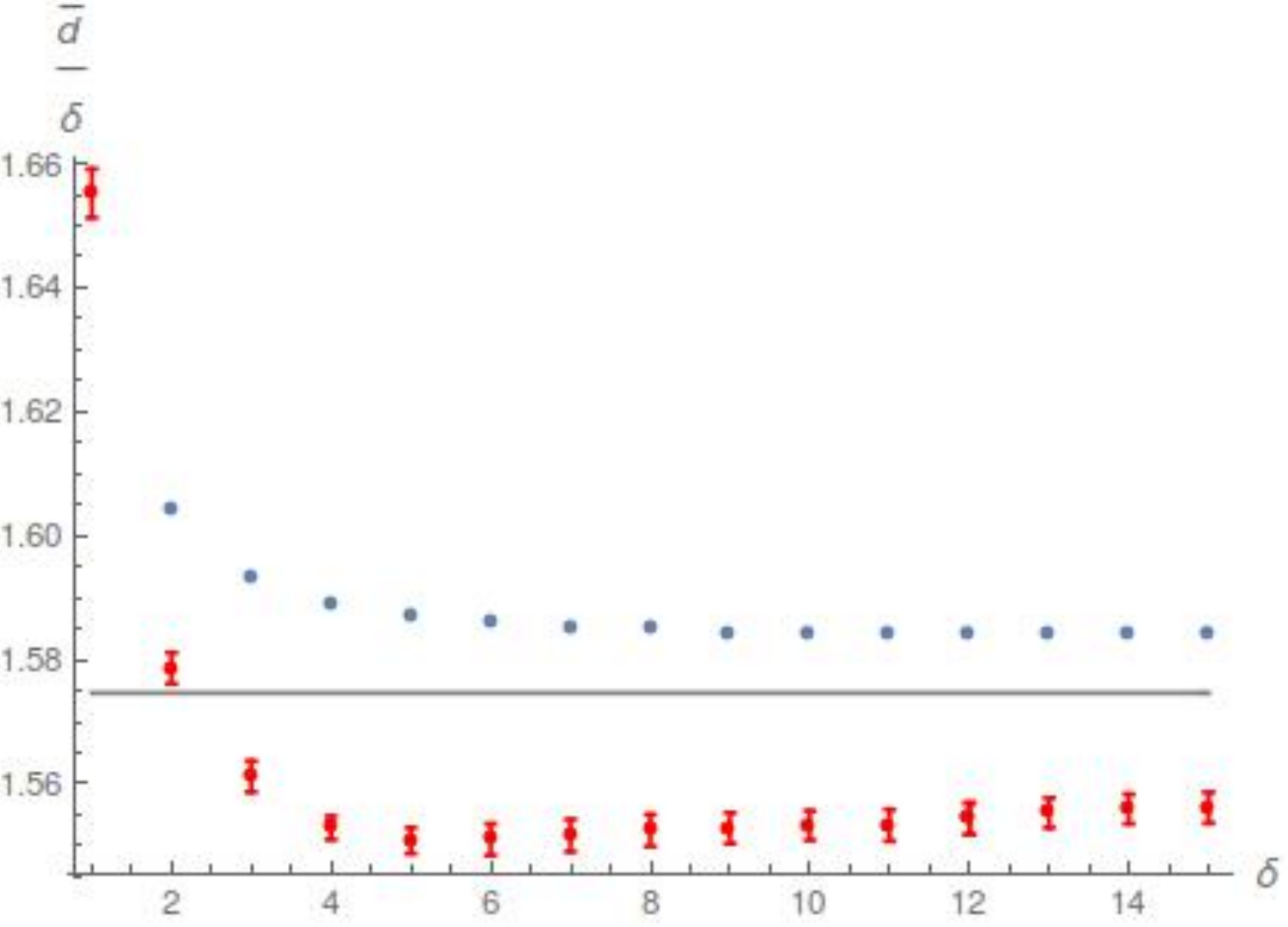}}}}
\caption{Normalized average sphere distance $\bar{d}/\delta$ as a function of the scale $\delta$, 
measured on random triangulations modelled on flat space (red data
points with error bars). For comparison, we have included the corresponding data for the flat hexagonal lattice (blue dots) of 
Fig.\ \ref{fig:2dregularav} and the horizontal line marking the constant value of continuum flat space (grey). 
}
\label{fig:curvmeasflat}
\end{figure}

Our measurements of the normalized average sphere distance $\bar{d}/\delta$ on the random triangulations at hand are shown in
Fig.\ \ref{fig:curvmeasflat}, where we have included the data for the regular hexagonal lattice and the constant continuum result for
flat space for comparison. The behaviour of the random triangulation is qualitatively similar to that of the hexagonal lattice: 
for small $\delta\!\geq\! 1$, $\bar{d}/\delta$ has initially a maximum, then decreases, and for $\delta\! \gtrsim\! 5$
settles to an approximately constant value, consistent with flat-space behaviour. Unlike what we saw for the regular flat lattices,
this value is now slightly below that for continuum flat space, and lies at approximately 1.55. The amplitude of 
the initial overshoot is in the same ballpark as those for the regular lattices (Fig.\ \ref{fig:2dregular}). From this
point of view, any nontrivial curvature that is present in the random triangulation on short scales is mixed with and
indistinguishable from the pure discretization effects of the flat lattices, as far as the quantum Ricci curvature is concerned.

To construct random triangulations modelled on non-flat spaces, we set without loss of generality the curvature radius of
the underlying sphere and hyperboloid to one, $\rho\! =\! 1$. Choosing different values of $d_{\rm min}$ for the Poisson
disc sampling then amounts to different degrees of fine-graining of the resulting triangulations with respect to this 
continuum reference scale.
A smaller $d_{\rm min}$ corresponds to a finer-grained triangulation and therefore to a smaller value of the local curvature.
After setting the edge lengths to 1, we expect to see these differences reflected in terms of lattice units. That is, 
we expect our measurements to be governed by an ``effective curvature radius" $\rho_{\rm eff}$ in lattice units, which 
is inversely proportional to $d_{\rm min}$. Moreover, by rescaling the results for random triangulations with different
fine-grainings in such a way that their effective curvature radii coincide, we expect their measurement data to fall on top of each other.

We have studied the situation in considerable detail for the case of the sphere, where we have worked with three distinct 
continuum cutoffs, $d_{\rm min}= 0.1,\, 0.05$ and $0.025$, and three different types of measurements from which 
an effective curvature radius can be extracted. Before embarking on these, we determined the vertex order distributions 
for the Delaunay triangulations of the sphere and found that for all three sizes considered they are
almost indistinguishable from that of flat space depicted in Fig.\ \ref{fig:vertexdist}.

\begin{figure}[t]
\centerline{\scalebox{0.5}{\rotatebox{0}{\includegraphics{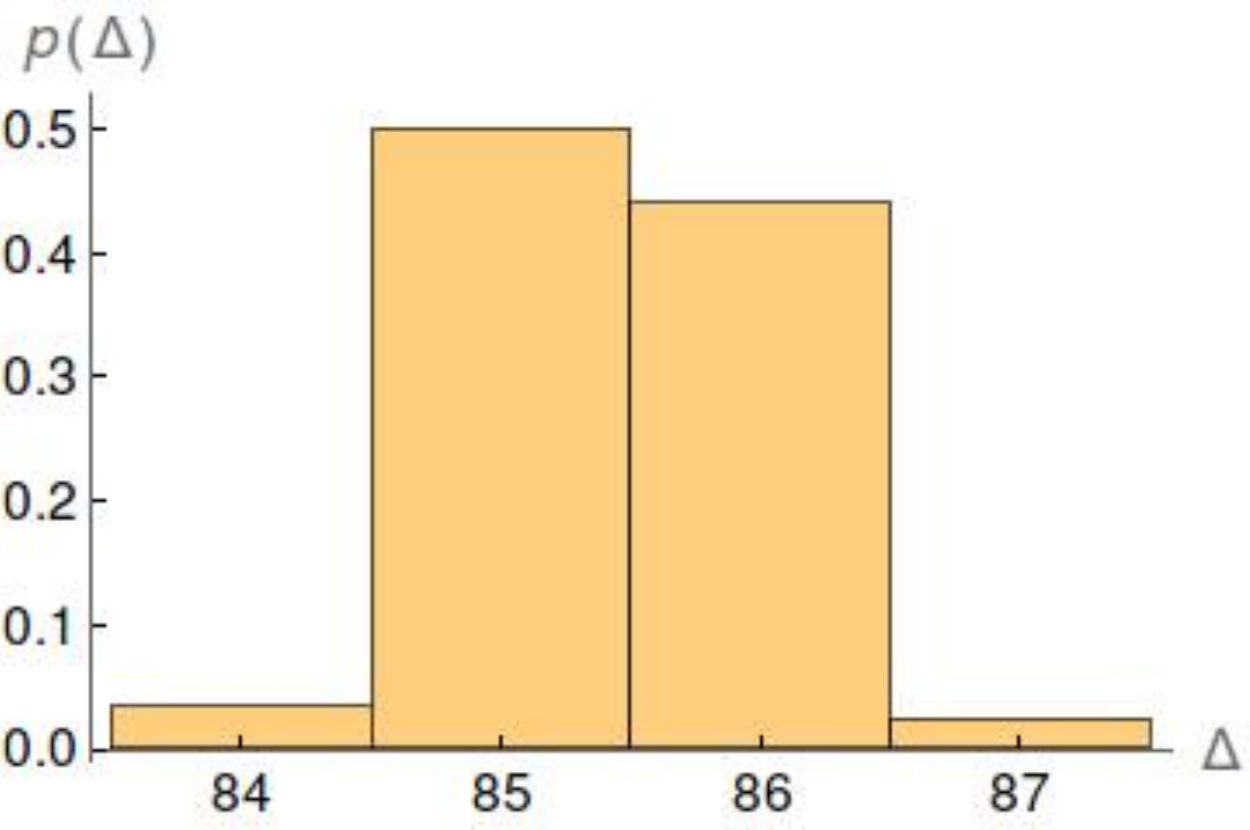}}}}
\caption{Distribution $p(\Delta)$ of diameters $\Delta$ of a sample spherical triangulation, generated
using $d_{\rm min}=0.025$. 
}
\label{fig:diamdistr}
\end{figure}

We then measured the distribution of diameters $\Delta$ of the triangulations obtained after setting $\ell=1$ for all edges,
a quantity defined in subsection \ref{measure:sec} above. In all
cases the distributions are very narrow, further supporting the closeness of the configurations to round continuum spheres. 
An example is shown in Fig.\ \ref{fig:diamdistr} for a triangulation constructed with $d_{\rm min}=0.025$. 
For $d_{\rm min}\! =\! 0.1,\, 0.05,\, 0.025$, the average diameters (averaged over ten configurations) were measured to be
$\Delta=21.5(1)$, 42.9(1) and 85.5(1), which after division by $\pi$ leads to the effective radii $\rho_{\rm eff}=6.84(4),\,
13.65(3)$ and 27.22(3) respectively. Note that (within measuring accuracy) 
subsequent values differ by a factor of 2, as one would expect for consistency. As we will see below, these values are
slightly, but systematically smaller (by about 7\%) than those extracted from circle and curvature scaling, which do agree mutually.
A possible explanation is that -- unlike the latter quantities -- the diameter by construction probes the largest scales of the
lattices, and therefore is subject to systematic finite-size effects.

\begin{figure}[t]
\begin{tabular}{ll}
\includegraphics[scale=0.55]{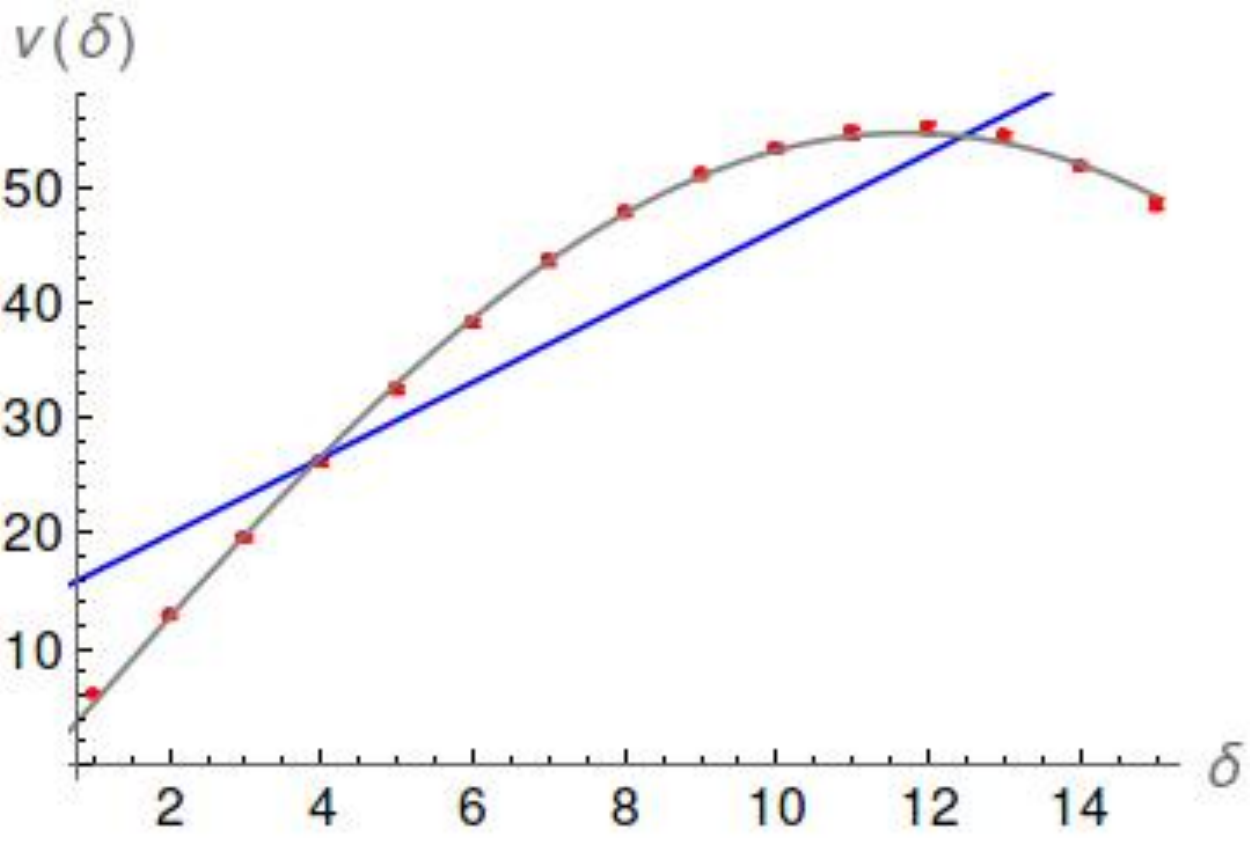}
&
\includegraphics[scale=0.55]{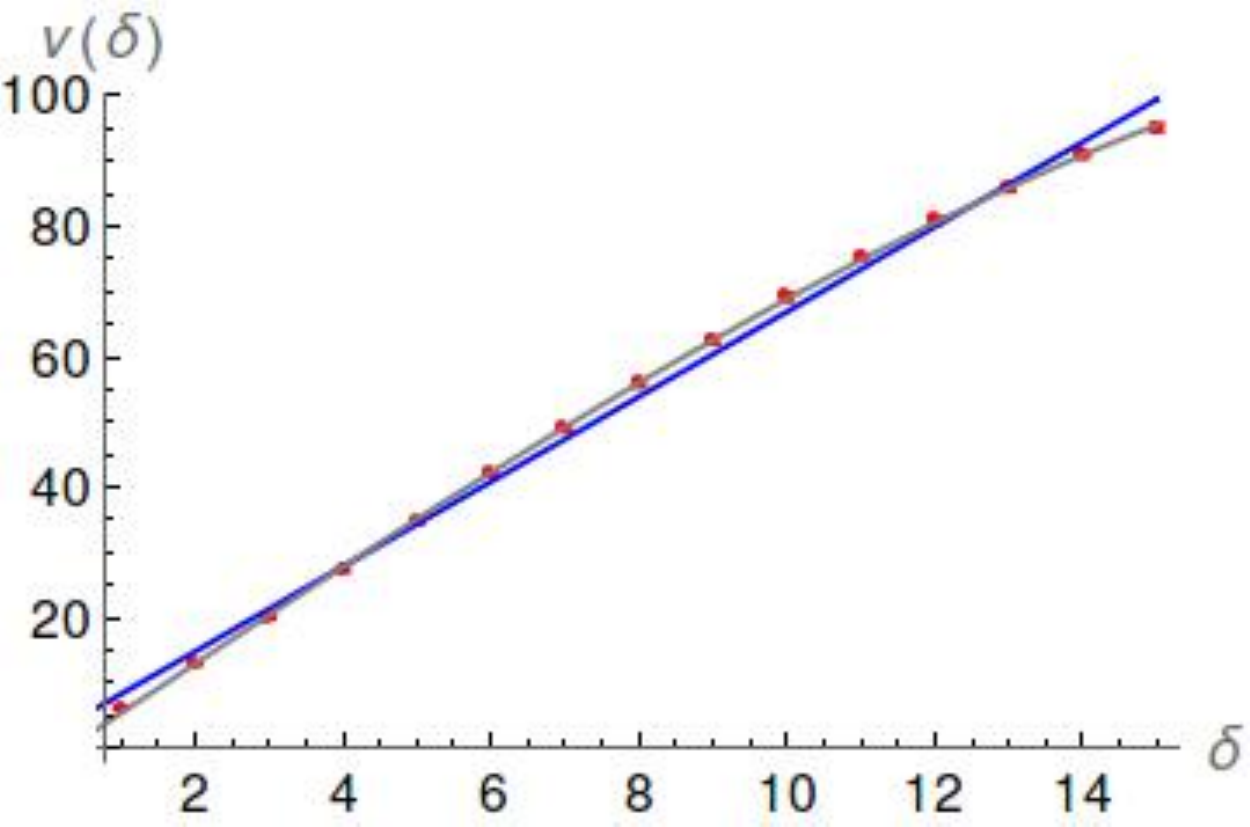}
\end{tabular}
\caption{The (averaged) size $\nu(\delta)$ of circles as a function of their radius $\delta$ on spherical triangulations,
for $d_{\rm min}\! =\! 0.1$ (left) and $d_{\rm min}\! =\! 0.05$ (right). We have included best fits to a
function of the form $c\rho_{\rm eff} \sin(\frac{\delta}{\rho_{\rm eff}}+s)$ (grey curves), and to a linear function (blue curves).
}
\label{fig:spherecurv}
\end{figure}

Next, we investigated the scaling of circle sizes $\nu (\delta)$ as a function of their geodesic radius $\delta$, and
compared them to the continuum formula $\nu (\delta)=2\pi\rho \sin(\frac{\delta}{\rho})$. This gives us another way of
extracting an effective curvature radius. However, in view of the analogous results for the flat case, we expect the overall
factor to deviate from $2\pi$. Furthermore, we have found that a (small) offset in $\delta$ improves the quality of the fits.
The need for such a shift may have to do with the fact that for topological reasons (by virtue of the Gauss-Bonnet theorem
for the two-sphere), the data is forced to
go through the point $\nu(1)=6$, resulting in a distortion for small $\delta$. The fitting function we have used is
\begin{equation}
\nu (\delta)= c\;\!\rho_{\rm eff}\, \sin\left( \frac{\delta}{\rho_{\rm eff}}+s\right) ,
\label{s-fit}
\end{equation}
for constants $c$ and $s$. 
Fig.\ \ref{fig:spherecurv} illustrates the situation for the two larger values of $d_{\rm min}$. 
In both cases, the sine function fits the data well. For comparison, 
we have also included linear fits to the data, but these are clearly inferior. 
\begin{table}[b]
  \begin{center}
  \begin{tabular}{c |  c c c  }
  $d_{\rm min}$ &  $s$ & $c$ & $\rho_{\rm eff}$ \\
  \hline
  0.1 &     $-4.1(7)\cdot 10^{-2}$ &  7.5(1) &   7.26(7) \\
  0.05 &     $-1.9(3)\cdot 10^{-2}$  & 7.6(4) &   15.6(5) \\
  0.025 &   $-6(2)\cdot 10^{-3}$  &   7.4(23)  &  47(17) \\
  \end{tabular} 
  \end{center}
  \caption{The parameters $s$, $c$ and $\rho_{\rm eff}$ obtained from fitting circle sizes to the functional form
$c\rho_{\rm eff} \sin(\frac{\delta}{\rho_{\rm eff}}\! +\! s)$, on spherical configurations of different sizes.}
   \label{fitdata}
  \end{table}

Obviously, within the limited range of $\delta$-values we are exploring, 
it becomes more difficult to distinguish between flat and curved space as the (effective) curvature radius
increases. This is illustrated by our last set of measurements, corresponding to $d_{\rm min}\! =\! 0.025$, where within
measuring accuracy the sine and linear functions fit the data about equally well. Not surprisingly, our estimate for the effective
curvature radius has very large error bars. Table \ref{fitdata} summarizes the values for the constants $s$ and $c$ and the
effective curvature radius $\rho_{\rm eff}$ obtained from best fits of $\nu (\delta)$, for the three different values of $d_{\rm min}$. 
 
The constant $c$ is approximately constant, 
which is consistent with having a single, overall scale factor for the length of geodesic circles, compared to the continuum, independent 
of sphere size. The values lie within one standard deviation from the corresponding value 7.48(5) 
we found in the flat case. 

\begin{figure}[t]
\centerline{\scalebox{0.7}{\rotatebox{0}{\includegraphics{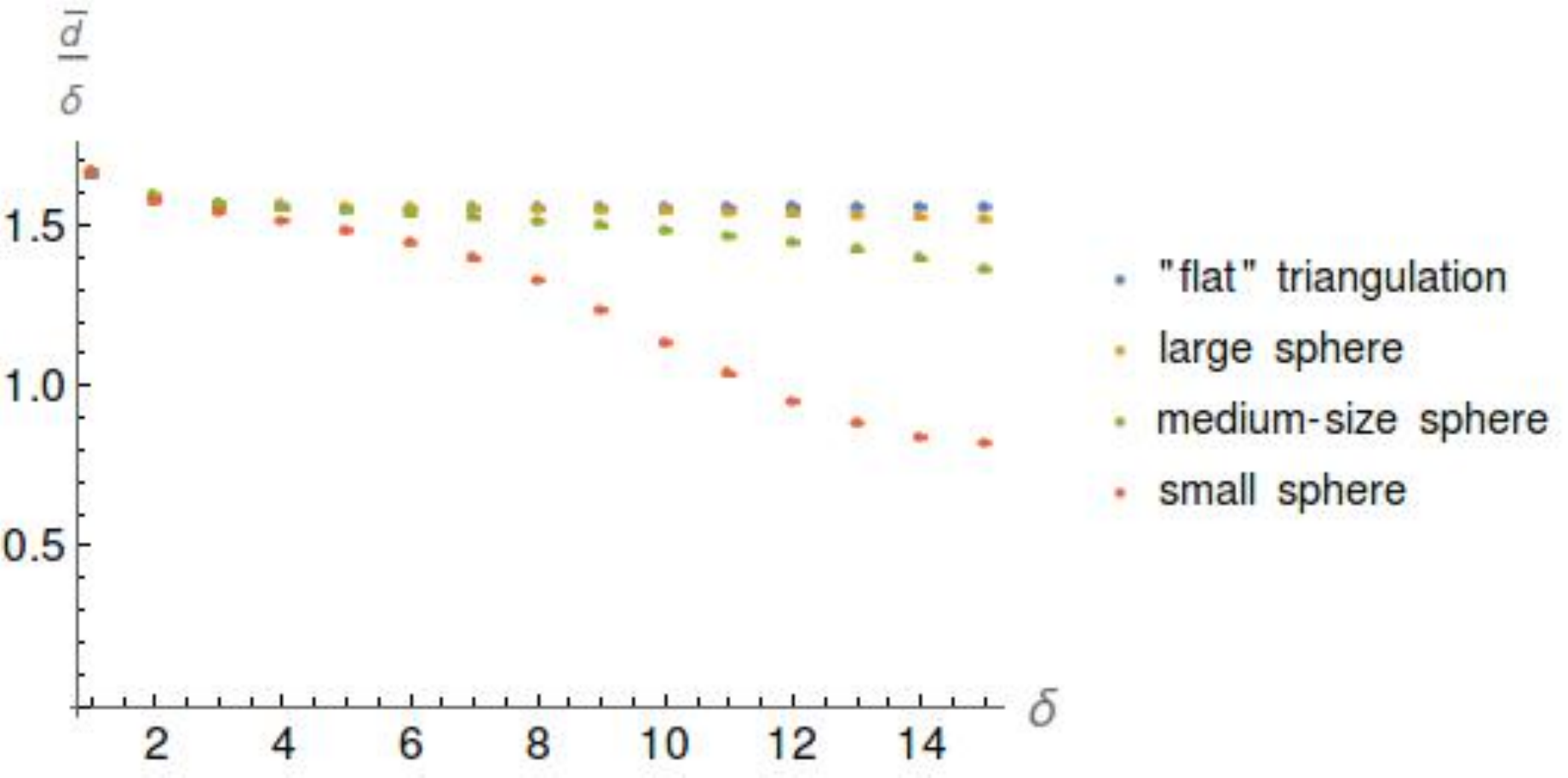}}}}
\caption{Normalized average sphere distance $\bar{d}/\delta$ as a function of the scale $\delta$, 
measured on random triangulations modelled on continuum spheres of three different sizes.
From top to bottom: ``flat" random triangulation measured previously (for reference); large sphere
($d_{\rm min}\! =\! 0.025$), medium-size sphere ($d_{\rm min}\! =\! 0.05$), and small sphere ($d_{\rm min}\! =\! 0.1$).
}
\label{fig:allspheres}
\end{figure}

Turning now to the measurements of the average sphere distance $\bar{d}(S^\delta_p,S^\delta_{p'})$, 
Fig.\ \ref{fig:allspheres} shows averaged values for the normalized quantity $\bar{d}/\delta$ for the three spherical
triangulations, including the data for the flat random triangulation (from Fig.\ \ref{fig:curvmeasflat}) for comparison. 
Qualitatively, the behaviour is as one would expect from the continuum: when moving to larger distances $\delta$, the ratio
$\bar{d}/\delta$ for the spheres goes to smaller values. The deviation from the horizontal flat-case line is largest for the
smallest sphere, whose positive curvature is largest. For the largest sphere, the deviation from the flat case 
can be seen quite clearly for the largest measured values of $\delta$, unlike the data from the circle scaling that did not
allow us to distinguish between the two cases. 

In order to make a quantitative comparison with the continuum, we would like to fit the data to curves 
of $\bar{d}/\delta$ for continuum spheres.
However, we need to account for the observed difference
in the constant $c_q$ of eq.\ (\ref{qric}) between the continuum geometries on the one hand and regular lattices and 
triangulations modelled on constantly curved spaces on the other. This requires an additional rescaling of
$\bar{d}/\delta$. There are two simple ways of achieving this, by
applying either a multiplicative scaling or a constant, additive shift to $\bar{d}/\delta$. As we will see, both types of
fit lead to similar results. To fix the additional matching parameter between continuum and discrete data, we require all curves to
go through the data reference point at $\delta\! =\! 5$. It is natural to anchor the curves at 
this point, because it is the approximate location on the $\delta$-axis where lattice artefacts seize to be significant.

In either case one is left with a one-parameter set of continuum curves, corresponding to different values of $\rho$.  
Among this set, we looked for the curve which best fitted our data, using a $\chi^2$-fit for data points in the interval 
$\delta\in[6,15]$. The smaller the sphere, the better is the quality of the fit. 

\begin{figure}[t]
\centerline{\scalebox{0.7}{\rotatebox{0}{\includegraphics{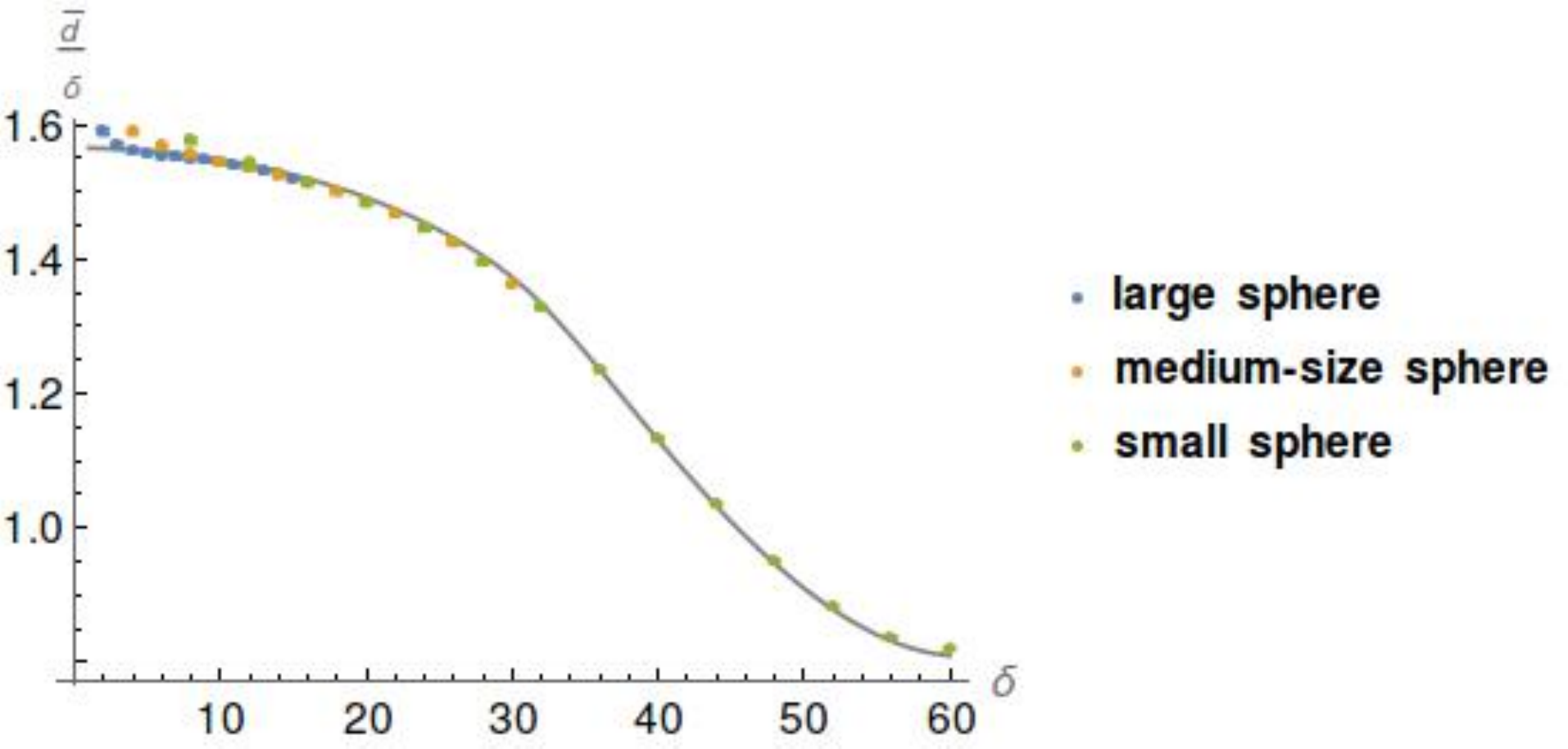}}}}
\caption{Measurements of $\bar{d}/\delta$ and best fit (using a multiplicative shift)
to the corresponding data of a two-dimensional continuum sphere, 
for the combined and rescaled data of all three spheres. Error bars are smaller
than dot sizes.
}
\label{fig:spherefits}
\end{figure}
\begin{table}[b]
\begin{center}
\begin{tabular}{c |  c c  }
$d_{\rm min}$ &  $\rho_{\rm eff}$, additive fit & $\rho_{\rm eff}$, multiplicative fit \\
\hline
0.1 & 7.41(12)  & 7.35(9) \\
0.05 & 14.31(24) & 14.27(21) \\
0.025 & 29.1(10) &  29.0(13)  \\
\end{tabular} 
\end{center}
\caption{Effective curvature radius $\rho_{\rm eff}$ of triangulations modelled on spheres, 
extracted from measuring the normalized average sphere distance, and fitting
to continuum spheres, using an additive or multiplicative shift of the data, as described in the text.}
\label{table2}
\end{table}
The results for the effective curvature radius extracted from fitting to continuum spheres are collected 
in Table \ref{table2}. We see that the two different types of fit lead to essentially identical results.  
Rescaling and combining the data for all three spherical configurations illustrates well that they can be
fitted to a single continuum curve, modulo short-scale deviations (Fig.\ \ref{fig:spherefits}), supporting the
existence of a universal underlying function $f(\delta/\rho)$. 

To combine the data and obtain the joint curve, 
we first multiplied the $\delta$-values of the data set for $d_{\rm min}\! =\! 0.1$ by a factor 4, and
that of $d_{\rm min}\! =\! 0.05$ by a factor 2, bringing them to the linear scale of the largest sphere.
The fit was obtained by considering the set of continuum curves going through the data point with $\delta\! =\! 5$ of the 
largest sphere and subsequently doing a $\chi^2$-fit involving the 10 data points for the largest $\delta$-values for each
of the three spheres, i.e. a total of 30 data points. 
The curvature radius associated with the combined curve is $\rho\! =\! 29.0(3)$,
corresponding to $\rho\! =\! 14.55$ for the medium-sized sphere and $\rho\! =\! 7.27$ for the small sphere, 
in very good agreement with
the effective curvature radii we extracted from individual spheres and from measuring circle volumes.  
However, it is worth noting that obtaining the curvature radius from the prescription for quantum Ricci curvature
for the same size of triangulation seems to give better results than obtaining it through 
circle scaling, despite the fact that the latter uses three instead of two fitting parameters.

Lastly, we report on the curvature analysis of the configurations obtained from the Delaunay triangulations on the two-dimensional 
hyperboloid. We performed measurements on ten independent configurations, which we constructed using $d_{\rm min}\! =\! 0.04$.
We had to restrict the $\delta$-range to $\delta\leq 13$ to avoid coming too close to the boundary of the
triangulations. The distribution of the interior vertex order resembles closely that of the flat and spherical cases. 
\begin{figure}[t]
\centerline{\scalebox{0.7}{\rotatebox{0}{\includegraphics{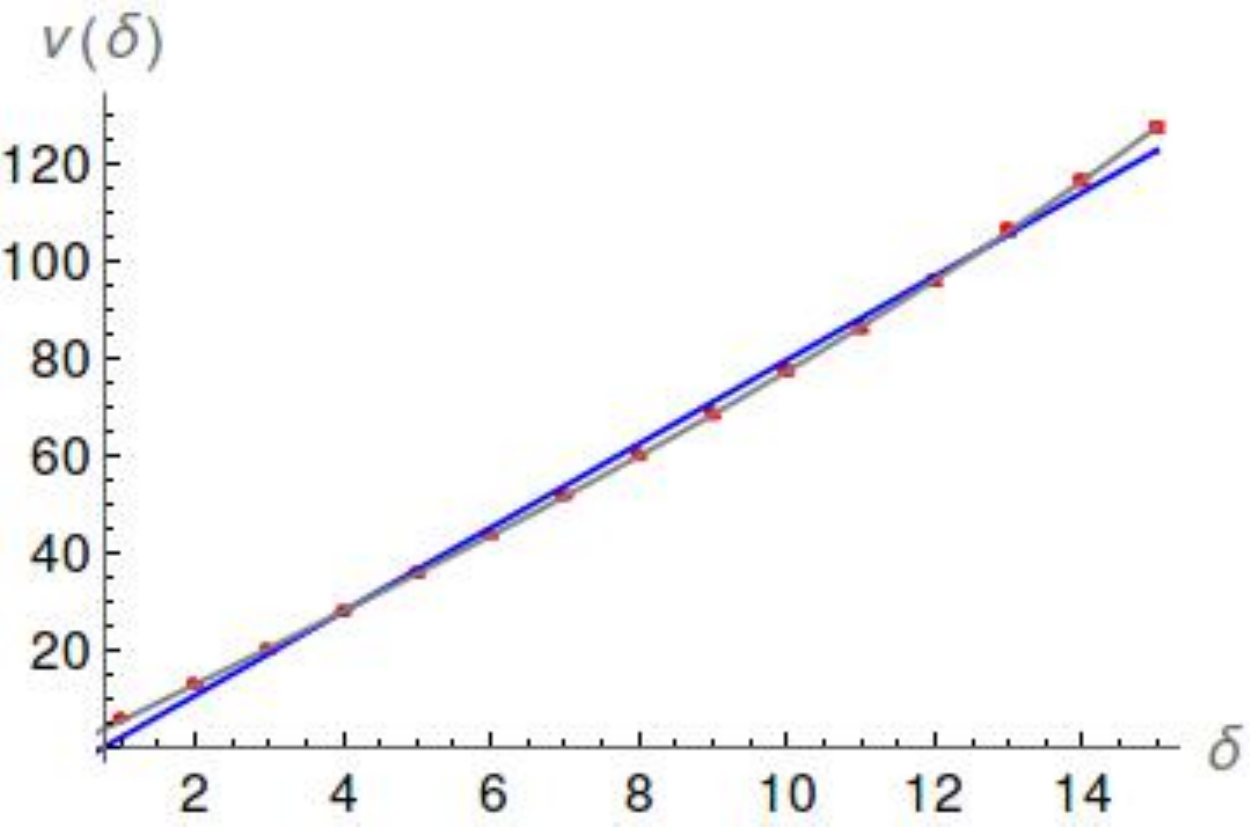}}}}
\caption{The (averaged) size $\nu(\delta)$ of circles as a function of their radius $\delta$ on 
geometries obtained by setting the edge lengths of Delaunay triangulations on a hyperboloid to unity,
including best fits to a
function of the form $\tilde{c}\rho_{\rm eff} \sinh(\frac{\delta}{\rho_{\rm eff}}+\tilde{s})$ (grey curve), 
and to a linear function (blue curve).
}
\label{fig:hypercircle}
\end{figure}
Next, we measured circle sizes $\nu(\delta)$ as a function of their radius $\delta$.
Following what we did in the spherical case, we used a three-parameter fit to extract an effective 
curvature radius $\rho_{\rm eff}$.
Substituting the sine by a hyperbolic sine function, we chose as a fitting function
\begin{equation}
\nu (\delta) = \tilde{c}\;\! \rho_{\rm eff}\, \sinh\left( \frac{\delta}{\rho_{\rm eff}}+\tilde{s}\right).
\label{h-fit}
\end{equation}
The continuum scaling would correspond to the special case
$\nu (\delta)\! =\! 2\pi \rho \sinh(\frac{\delta}{\rho})$.
From a best fit, we have determined the three parameters as $\tilde{s}=-1.69(17)\cdot 10^{-2}$, $\tilde{c}=7.4(3)$ 
and $\rho_{\rm eff}=15.0(5)$. Like in the spherical case, the shift $\tilde{s}$ is small. 
The measured data are 
plotted in Fig.\ \ref{fig:hypercircle}, together with the hyperbolic sine fit (\ref{h-fit}) and a linear fit through the origin for comparison.
The former clearly fits the data better, providing further evidence that our triangulations approximate constantly curved
continuum spaces also for negative curvature.

\begin{figure}[t]
\centerline{\scalebox{0.7}{\rotatebox{0}{\includegraphics{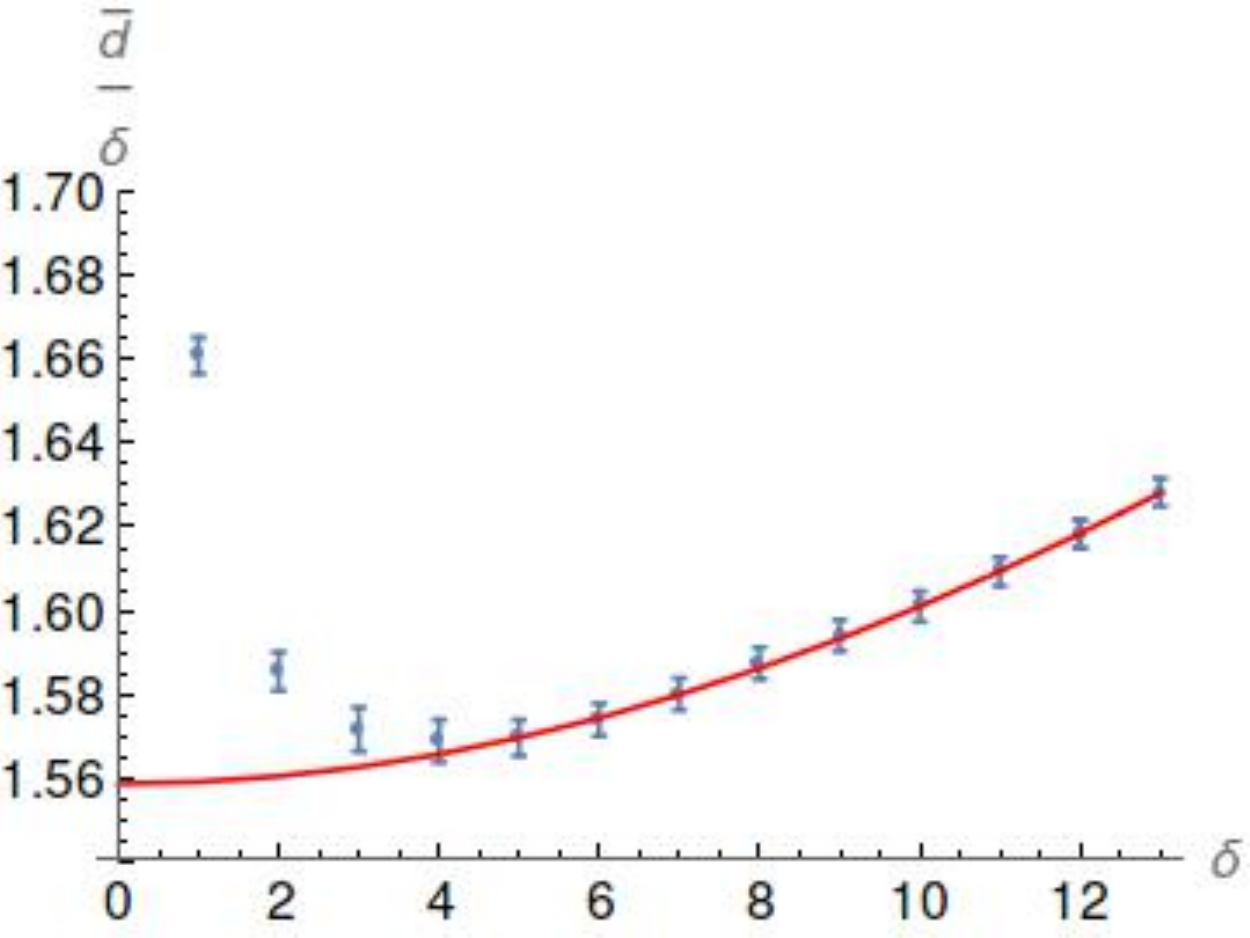}}}}
\caption{Normalized average sphere distance $\bar{d}/\delta$ as a function of the scale $\delta$, 
measured on random triangulations modelled on a continuum hyperboloid (blue), shown together
with a best fit of the corresponding curve in the continuum (red).
}
\label{fig:hyperboloid}
\end{figure}
The measurements of the normalized average sphere distance for the hyperbolic case are shown in Fig.\ \ref{fig:hyperboloid}.
We again performed two fits, a multiplicative and an additive rescaling of $\bar{d}/\delta$, combined
with the requirement that curves should pass through the data point at $\delta\! =\! 5$. 
Both result in a very good match with the data; the best fit for the additive rescaling is displayed in
Fig.\ \ref{fig:hyperboloid}. (It is barely distinguishable from the fit for multiplicative scaling.) 
As in previous measurements, there is
a short-distance regime where $\bar{d}/\delta$ exhibits an ``overshoot". 
From best matching for the data points $\delta\in[6,13]$, we determined the effective curvature radius as $\rho_{\rm eff}=18.0(3)$
for the additive fit and $\rho_{\rm eff}= 17.9(4)$ for the multiplicative fit. Both are in excellent agreement with
each other, but not with the value we extracted from the circle scaling. 

Comparing with the data for the smallest sphere,
and using the fact that we expect the product $d_{\rm min}\!\cdot\! \rho_{\rm eff}$ to be approximately constant,
one would expect the effective curvature radius in the hyperbolic case 
to lie in the interval $[18.0,18.5]$. While the data coming from measuring the quantum Ricci curvature 
are perfectly compatible with this estimate, the data from the circle scaling are off by six standard deviations.
The only plausible explanation we have at this stage is that the hyperbolic case suffers from finite-size effects, due
to the exponential growth of the volume with the radius, which for the volumes we are considering affect the
circle scaling, but apparently not the average sphere distances. This question can be settled by going to larger 
lattices, which is beyond the scope of our present work. However, the encouraging message is that 
in comparison, the measurement of the quantum Ricci curvature again appears to be more robust.

\section{Summary, conclusions and outlook}
\label{conclusion:sec}

In this paper, we have defined a new way of quantifying the curvature properties of metric spaces
in terms of ``quantum Ricci curvature".
Our starting point was the known observation that on smooth spaces the distance between two
spheres in general differs from the distance between their centres in a way that depends on the Ricci curvature. 
Building on this observation, we
constructed a curvature observable that is scalable and straightforward to compute, as we have demonstrated in many 
explicit examples. We defined the quantum Ricci curvature initially on purely classical, Riemannian manifolds, using a
generalized notion of distance between spheres, based on averaging over both spheres. 
This replaces the transportation distance (sometimes also called Wasserstein distance) used in the Ollivier curvature
\cite{ollivier, ollivier1}. A main motivation was computability, especially in view of the fact that we want to evaluate
the curvature also on large scales. 

One could investigate the properties of quantum Ricci curvature in the classical continuum context in greater
detail, but the prime aim of our current study was to show its feasibility in generalized, non-smooth settings, preparing
the ground for its application in fully fledged quantum gravity. We limited our continuum analysis to the evaluation of the 
quantum Ricci curvature on two-dimensional spaces of constant curvature,
which gave us a first quantitative grasp of the large-scale behaviour of this quantity.
Note that on a two-dimensional Riemannian manifold the local Ricci curvature $Ric(v,v)$, for any vector $v$, coincides (up to a factor of 2)
with the Ricci scalar.\footnote{This is no longer true in higher dimensions, where the evaluation of the normalized average
sphere distance for infinitesimal $\delta\! =\! \epsilon$ at order $\delta^3$ yields a linear combination of the Ricci curvature and
the Ricci scalar \cite{qrc2}.}
The characteristic behaviour of the normalized average sphere distance for positive, zero and negative
curvature shown in Fig.\ \ref{fig:diagnorm} also served as a benchmark for our subsequent curvature measurements
on non-smooth spaces.

We described in Sec.\ \ref{cqc:sec} the challenge of defining a meaningful notion of curvature on non-smooth metric
spaces, which in general lack a differentiable structure and the tensorial quantities that go with it. This raises the question
of how a genuine tensor like the Ricci curvature $R(v,v)$ associated with a vector $v$ can be represented. 
The analogue of a vector $v$ of length $\delta$ in our implementation of the quantum Ricci curvature is given by
a pair of overlapping spheres or balls of radius $\delta$. When $\delta$ is an integer, like in the piecewise
flat spaces we considered, the smallest value where the quantum Ricci curvature can be evaluated is $\delta\! =\! 1$,
which is why we call it a ``quasi-local" quantity. 

Our analysis of the quantum Ricci curvature on piecewise flat spaces was motivated directly by the nonperturbative
quantum theory formulated in terms of causal dynamical triangulations. As already emphasized in the introduction, 
the triangular building blocks in that case play the role of a short-distance regulator: the space of all $D$-dimensional 
spacetimes -- the configuration space of the gravitational path integral -- is approximated by a space of simplicial manifolds 
whose building blocks are equilateral $D$-simplices of some fixed edge length $a$. Since the details of the chosen
regularization should not matter in the final continuum theory, physically interesting continuum limits 
$a\rightarrow 0$ should not depend on them at any scale, including the Planck scale. We discard measurements near
the cutoff $a$ as ``discretization artefacts", because they usually bear a strong imprint of these details.
In this respect our perspective on generalized Ricci curvature is different from that
frequently taken in discrete mathematics and network theory, where the discrete, short-scale
structure in itself is the primary focus of interest. This is also the case in recent implementations of
Ricci curvature \`a la Ollivier in attempts to construct a theory of quantum gravity from specific statistical ensembles of
random graphs or networks \cite{trugenberger} (see also \cite{bianconi} for related ideas).

With the large-scale perspective in mind, we first studied the behaviour of the quantum Ricci curvature on regular flat
lattices in two and three dimensions. These structures are ``flat", in the sense that they can be imbedded in flat Euclidean
space, from which they inherit their (unit) edge length assignments.
We can treat these lattices as piecewise flat structures and work with the discrete geodesic link distance to compute 
lengths and geodesic spheres, thus providing a first test of the quantum Ricci curvature in a discretized setting.

All regular lattices we investigated display some common characteristics. They have a short-distance regime where the 
normalized average sphere distance $\bar{d}/\delta$ starts out at some maximum value for $\delta\! =\! 1$ and
then decreases rapidly until about $\delta\! =\! 5$, where the $\bar{d}/\delta$-curve enters its flat regime. 
The presence (in our interpretation) of lattice artefacts below $\delta\! =\! 5$
means that we should not consider the limit $\delta\rightarrow 0$
to extract the constant $c_q$ of relation (\ref{qric}), as we did in the continuum, but rather evaluate 
$\bar{d}/\delta$ at $\delta\! =\! 5$, or elsewhere in the constant region. 
Following this logic, we found
that the value of $c_q$ differs from the corresponding value in the smooth case, and also depends on the lattice type.

Since the regular lattices can be thought of as simple discretizations of flat space, one would expect them to
behave like flat spaces in the continuum sense on scales that are sufficiently large in terms of lattice units.
The corresponding lattice- and discretization-independent analogue of this behaviour appears to be the
vanishing of the quantum Ricci curvature, $K_q\! =\! 0$ (or, equivalently, the constancy of the quantity $\bar{d}/\delta$), 
providing further justification for the ansatz (\ref{qric}).

The equilateral random triangulations we investigated next probe different properties of the quantum Ricci curvature.
We constructed these triangulations with the intention of having them resemble constantly curved continuum spaces
on large scales, while introducing random curvature fluctuations on small scales. 
It is not a foregone conclusion that the local small-scale curvature will ``average out" on coarse-grained scales
with respect to any measure of curvature, 
but this is exactly what we observed when evaluating the quantum Ricci curvature as a function of the scale $\delta$.

For the triangulated spaces modelled on Delaunay triangulations of flat space, the results for the normalized average sphere
distance resembled closely those of the regular flat lattices. Within measuring accuracy, the $\bar{d}/\delta$-curve is
flat for distances $\delta \gtrsim 5$, signalling a vanishing of the quantum Ricci curvature. For smaller $\delta$, 
applying formula (\ref{qric}), the quantum Ricci curvature is nominally negative, but since we have already identified 
this region as dominated by lattice artefacts, this statement has little physical significance. The same is true for
the short-distance behaviour of the equilateral random triangulations modelled on Delaunay triangulations of curved spaces. 
The measurements of the normalized average sphere distances for $\delta\!\geq\! 5$ in these cases could be matched
well to the corresponding continuum curves for spheres and hyperboloids. After performing a single shift in $\bar{d}/\delta$ 
to account for the a priori unknown $c_q$-value of a given type of piecewise flat space, we extracted  
effective curvature radii from a best matching to the continuum curves. 
All results were consistent with each other (e.g. for different sphere sizes) and consistent with the behaviour 
of the constantly curved continuum spaces they were meant to approximate in the first place. We also noted in
passing that extracting the effective curvature radius from measuring the quantum Ricci curvature seems to give
more accurate results than obtaining it from the scaling of sphere sizes.

To summarize, our analytical and numerical investigations of the novel quantum Ricci curvature on ``nice" 
equilateral triangulations of moderate size, mostly in two dimensions, have demonstrated that it can be implemented
and measured in a straightforward way. Lattice artefacts are confined to a scale of about five lattice spacings,
above which the behaviour of the quantum Ricci curvature conforms with continuum expectations. 
In other words, away from the cutoff scale it is sensitive to neither lattice discretization effects nor
the local curvature defects we introduced by removing the link length information from the Delaunay
triangulations. In our view, the observed robustness of the quantum Ricci curvature has to do with the
fact that the underlying normalized average sphere distance $\bar{d}/\delta$ is a dimensionless quotient of 
two quantities of the same kind, namely, an average distance and a distance, which will be affected by lattice
discretization effects in a similar way.

These promising results pave the way for an evaluation of the quantum Ricci curvature on a nonperturbative
quantum ensemble of spacetimes, like that of Causal Dynamical Triangulations. Of course, to obtain a
proper quantum observable, we must perform a suitable average over spacetime points. This will then in turn
be evaluated in the sense of eigenvalues, that is, by averaging over the spacetime
configurations in the ensemble. Our implementation of quantum Ricci curvature in two-dimensional
quantum gravity in terms of dynamical triangulations demonstrates
that such a procedure is feasible and meaningful, even in a situation where the underlying geometric configurations 
are very far removed from smooth classical spaces \cite{qrc2}. The results obtained in this case further underline the
robustness and good behaviour under averaging of the quantum Ricci curvature we found in the work presented here.

%\vspace{0.3cm}

\subsection*{Acknowledgments} 
This work was partly supported by the research program
``Quantum gravity and the search for quantum spacetime" of the Foundation for Fundamental Research 
on Matter (FOM, now defunct), financially supported by the Netherlands Organisation for Scientific Research (NWO).
It was also supported in part by Perimeter Institute for Theoretical Physics. Research at Perimeter Institute is supported 
by the Government of Canada through Industry Canada and by the Province of Ontario through the Ministry of Economic 
Development and Innovation.
\vspace{0.3cm}

%\newpage

\end{document}